\documentclass[structabstract]{aa}  
\usepackage{graphicx}
\begin{document}
\title{The size-luminosity relation at z=7 in CANDELS and
its implication on reionization}

   \author{A. Grazian \inst{1}
          \and
          M. Castellano \inst{1}
          \and
          A. Fontana \inst{1}
          \and
          L. Pentericci \inst{1}
          \and
          J. S. Dunlop \inst{2}
          \and
          R. J. McLure \inst{2}
          \and
          A. M. Koekemoer \inst{3}
          \and
          M. E. Dickinson \inst{4}
          \and
          S. M. Faber \inst{5}
          \and
          H. C. Ferguson \inst{3}
          \and
          A. Galametz \inst{1}
          \and
          M. Giavalisco \inst{6}
	  \and
	  N. A. Grogin \inst{3}
          \and
          N. P. Hathi \inst{7}
	  \and
	  D. D. Kocevski \inst{5}
	  \and
	  K. Lai \inst{5}
          \and
          J. A. Newman \inst{8}
          \and
          E. Vanzella \inst{9}
          }

   \offprints{A. Grazian, \email{andrea.grazian@oa-roma.inaf.it}}

\institute{INAF - Osservatorio Astronomico di Roma, Via Frascati 33,
I--00040, Monteporzio, Italy
\and SUPA, Institute for Astronomy, University of Edinburgh,
Royal Observatory, Edinburgh EH9 3HJ, UK
\and Space Telescope Science Institute, 3700 San Martin Drive, Baltimore,
MD 21218, USA
\and NOAO, 950 N. Cherry Avenue, Tucson, AZ 85719, USA
\and UCO/Lick Observatory, University of California, 1156 High
Street, Santa Cruz CA, 95064, USA
\and Department of Astronomy, University of Massachusetts, 710 North
Pleasant Street, Amherst, MA 01003, USA
\and Carnegie Observatories, 813 Santa Barbara Street, Pasadena, CA 91101, USA
\and Department of Physics and Astronomy, University of Pittsburgh, Pittsburgh,
PA 15260, USA
\and INAF - Osservatorio Astronomico di Trieste, Via G.B. Tiepolo 11,
I--34131, Trieste, Italy
}

   \date{Received May 24, 2012; accepted August 25, 2012}

   \authorrunning{Grazian et al.}
   \titlerunning{The size-luminosity relation at $z\sim 7$}

 
  \abstract
   {
The exploration of the relation between galaxy sizes and other physical
parameters (luminosity, mass, star formation rate) has provided important clues
for understanding galaxy formation, but such exploration has until recently
been limited to intermediate redshift objects.
   }
   {
We use the currently available CANDELS Deep+Wide surveys in the GOODS-South,
UDS and EGS fields, complemented by data from the HUDF09
program, to address the relation between size and luminosity at $z\sim 7$.
   }
   {
The six different fields used for this study are
characterized by a wide combination of depth and areal coverage, well
suited for reducing the biases the observed size-magnitude plane. From
these fields, we select 153 z-band dropout galaxies.
Detailed simulations
have been carried out for each of these six fields, inserting simulated
galaxies at different magnitudes and half light radius in the two dimensional
images for all the HST bands available and recovering them as carried out
for the real galaxies. These simulations allow us to derive precisely the
completeness as a function of size and magnitude and to quantify
measurements errors/biases, under the assumption that the 2-D profile of
z=7 galaxies is well represented by an exponential disk function.
   }
   {
We find in a rather robust way that the half light radius
distribution function of $z\sim 7$ galaxies fainter than $J=26.6$ is
peaked at $\le 0.1$ arcsec (or equivalently 0.5 kpc proper), while at
brighter magnitudes high-z galaxies are typically larger than
$\sim$0.15 arcsec. We also find a well defined
size-luminosity relation, $Rh\propto L^{1/2}$. We compute the
Luminosity Function in the HUDF and P12HUDF fields, finding large
spatial variation on the number density of faint galaxies.
Adopting the size distribution and the size-luminosity relation found
for faint galaxies at z=7, we derive a mean slope of $-1.7\pm 0.1$
for the luminosity function of LBGs at this redshift.
   }
   {
Using this LF, we find that the number of ionizing photons emitted from galaxies
at $z \sim 7$  cannot keep the
Universe re-ionized if the IGM is clumpy ($C_{HII}\ge 3$) and the
Lyman continuum escape fraction of high-z LBGs is relatively low
($f_{esc}\le 0.3$).
If these results are confirmed and strengthened by future CANDELS data,
in particular by the forthcoming deep observations in GOODS-South and North
and the wide field COSMOS, we can put severe limits to the
role of galaxies in the reionization of the Universe.
   }


   \maketitle
%

\section{Introduction}

The advent of the WFC3 instrument onboard HST has opened a new window
for the study of galaxy shape, size, and morphology up to
very high redshifts (\cite{windhorst}). The combination of large
area, fine resolution, and NIR wavelengths achieved with this powerful
instrument allows us to study galaxies in the UV rest frame at $z\sim
7-8$ with impressive accuracy.

An important parameter that is useful for constraining different models of
galaxy formation and evolution is galaxy size, measured through its half
light radius (hereafter $Rh$). This quantity gives us an indication of the
dynamical state of the galaxy itself and the effects of feedback,
minor/major merging, inflows and outflows. In particular, the
relation between size and luminosity, or other physical properties
(stellar mass, dust extinction, etc.), can give insight into the
detailed galactic assembly processes.

With the advent of large surveys, like the Sloan Digital Sky Survey
(SDSS, \cite{sdss}) it has been possible to study the physical
properties of local galaxies with great accuracy. Present-day
galaxies show a clear correlation between size and stellar mass, with
the most massive galaxies having the largest half-light radii
(\cite{shen03,gw08}). This mass-size correlation has been found both for
elliptical and spiral galaxies. In particular, \cite{barden05} pointed
out that the same relation holds up to $z\sim 1$ and its normalization
is unchanged with respect to local disk galaxies at least for stellar masses
$M\ge 10^{10}M_{\odot}$. Recently, \cite{mosleh12} extended the evolution of
the stellar mass-size relation for star-forming galaxies till $z\sim 7$,
finding that the typical size of LBGs increases toward lower redshifts,
in agreement with previous measurements at low-z.

Star Forming galaxies at $z\sim 2-3$, instead, have been extensively
studied using ground based spectroscopy, HST imaging and IFU
observations. Lyman Break Galaxies (LBGs, \cite{sh93,madau}) at $z\sim
2-3$ show a stellar mass-radius relation already established
(\cite{nagy,law}). Similarly, Lyman-$\alpha$ emitters (LAEs) at the
same redshifts present a correlation between their sizes and other
physical properties, such as stellar mass, star formation rate (SFR),
Spectral Energy Distribution (SED) or dust extinction (\cite{bond}),
with larger galaxies having higher stellar masses, higher dust
extinction, and higher SFR. The half light radius of these LAEs at
$z=2$, however, is not correlated to the EW in Lyman-$\alpha$. The
stellar mass-radius relation evolves in redshift as $(1+z)^{-1}$, in
a manner consistent with the size evolution found by \cite{bouwens04} and
\cite{ferguson04} for LBGs at $z\ge 2-5$ and by \cite{hathi08} at $z\sim 5-6$.

At $z\sim 6$, \cite{sizez6} found a well-defined correlation between measured
size and observed magnitudes for 332 photometrically selected LBG candidates:
this indicates that a size-luminosity relation could be still in place
at high-z. At magnitudes fainter than $z_{\rm AB}\sim 28$ there is a clear lack of
galaxies larger than 0.2 arcsec, but at such faint levels, the effect of surface
brightness dimming is limiting the completeness of large galaxies.
At the same redshift, \cite{Dow2007} found that all the Lyman-$\alpha$
emitters are more compact than average relative to the observed size-magnitude
relation of the large i-dropout sample of \cite{sizez6}.

The evolution of galaxy size with redshift and luminosity also has
important implications for the faint end of the Luminosity Function
(LF) and the role of low-luminosity galaxies in 
the reionization of the Universe. One of the main
motivations of this work is to answer to the questions raised in
\cite{grazian11}. In our previous work we have explored 
different distributions for the half light radius of $z=7$
galaxies. One of the clearest results is that the LF is quite steep
($\alpha \sim -2$) if faint galaxies are extended ($Rh\sim 0.2-0.3
arcsec$ at $J=28-29$) while it turns out to be similar to lower-z LFs
($\alpha \sim -1.7$) if objects become smaller at relatively faint
magnitudes. This is simply due to the corrections for incompleteness at the
faint end of the LF, which are more severe for large and extended galaxies.
A steep LF has deep implications for the number of ionizing
photons produced by galaxies at $z\sim 7$: a typical LF with $\alpha
\sim -1.7$ provides enough light to maintain the reionization process
only assuming a large escape fraction of Lyman Continuum photons
($f_{esc}>20\%$), an IGM that is not clumpy ($C_{HII}<4-6$), and
extrapolating this steepness down to very faint flux levels
($M_{1500}=-10$). These constraints are valid under the assumptions of
a Salpeter IMF and ignoring the effects of PopIII stars or other exotic
sources of ionizing radiation.
On the other hand, if faint galaxies are extended,
then the resulting LF is quite steep ($\alpha \sim -2$) and galaxies
alone are able to keep the Universe reionized even for less extreme
combinations of escape fraction and clumpiness ($f_{esc}>5\%$ and
$C_{HII}<30$). Thus a detailed analysis on the typical sizes of
high-z galaxies and the relation between galaxy size and luminosity is
necessary to understand whether galaxies alone are the responsibles
for reionization. In \cite{grazian11} we did not provide a definitive
answer to these questions, which therefore motivated the investigation
here with a larger sample of galaxies at $z\sim 7$.
Of course, other hypotheses on the sources responsible for reionization at
such high-z are possible, like a top-heavy IMF, a large contribution from
PopIII stars, or other sources of ionizing photons i.e. high-z AGNs.

Throughout this paper, we will assume a ``concordance'' cosmology with
$H_0=70km~s^{-1}~Mpc^{-1}$, $\Omega_M=0.3$ and $\Omega_\Lambda=0.7$.
In this cosmological model, an angular dimension of 1 arcsec corresponds
to a physical dimension of 5.227  kpc (proper) at z=7.


\section{Data}

\subsection{The photometric sample}

We have analysed six different data sets observed with the NIR camera
of HST, the Wide Field Camera 3
(WFC3\footnote{http://www.stsci.edu/hst/wfc3}): the UKIDSS Ultra Deep
Survey (UDS, 227 sq. arcmin to J=26.7) and the Extended Groth Strip
(EGS, 110 sq. arcmin to J=26.7) from the CANDELS-Wide survey
(\cite{grogin11,koekemoer11}), the Early Release Science
(\cite{windhorst}) on the GOODS-S field (GOODS-ERS, 40 sq. arcmin. to
J=27.4), part of the GOODS-South Deep (GDS, 27 sq. arcmin. to J=27.8)
from the CANDELS-Deep survey (\cite{grogin11,koekemoer11}), the first
year observations of the Hubble Ultra Deep Field (HUDF, 4.7
sq. arcmin. to J=29.2) and its parallel field (P12HUDF, 4.7
sq. arcmin. to J=29.2) described in \cite{bouwens10c}. All together,
they allow us to cover a broad enough range of galaxy size and
luminosity to investigate their interrelationship.

\subsubsection{UDS}

The UDS program is part of the CANDELS/Wide survey and it was the first wide
field to be observed by the CANDELS observations in the NIR with WFC3.
It is centered on the UKIDSS Ultra Deep
Survey field (\cite{lawrence07})
and benefits from the ground based imaging with SUBARU in BVRiz bands
(\cite{sxds}), deep JHK imaging from UKIRT-WFCAM, deep CFHT U band, MIR
observations by Spitzer (SpUDS and SEDS programs),
and intense follow-up optical spectroscopy from
Gemini\footnote{http://mur.ps.uci.edu/$\sim$cooper/IMACS/home.html},
VLT-VIMOS, and VLT-FORS2 (\cite{smail2008,simpson}).
This field has been covered by HST in optical and NIR using a mosaic
grid of tiles and repeated over two epochs. During each epoch, each tile
has been observed for one orbit ($\sim$2000 seconds),
divided into two exposures in J125 (at a depth of 1/3 orbit) and two
exposures in H160 (at a depth of 2/3 orbit), together
with parallel exposures using ACS/WFC in V606 and
I814. Observations and data reduction for the UDS field are
described in detail in \cite{koekemoer11} and \cite{grogin11}.

The total area covered by the CANDELS/UDS is $\sim 227$ sq. arcmin.
down to J=26.7, and H=26.5 magnitudes at $5\sigma$ in a circular aperture of
$\sim 0.11$ $arcsec^2$ (corresponding to 2 times the FWHM of the NIR images).

\subsubsection{EGS}

The EGS (\cite{egs}) field is part of the CANDELS/Wide survey. It covers
a region of the sky that has been extensively studied by the All-wavelength
Extended Groth strip International Survey (AEGIS).
This field has been observed with ACS in the V606 and I814 bands and by
a number of ground based facilities from the U to the K band. These data have
been complemented by observations with the Spitzer Space Telescope in the
3.6-70$\mu$m range, in X-ray by Chandra, in FUV by GALEX and in radio by VLA.
The CANDELS/Wide observations used here are only the first half of the area
that will be covered at the completion of the survey, namely a rectangular
grid of $6.7\times 30.6$ sq. arcmin. The WFC3 observing strategy in the EGS
mirrors that adopted for the UDS, covering a total area of $\sim 110$ sq.
arcmin.
down to J=26.7, and H=26.5 magnitudes at $5\sigma$ in a circular aperture of
$\sim 0.11$ $arcsec^2$ (corresponding to 2 times the FWHM of the NIR images).
An overview of observations available for the EGS field is
described in detail in \cite{grogin11}.

\subsubsection{ERS}

The GOODS-ERS WFC3/IR observations comprised 60 HST orbits consisting of
10 contiguous pointings in the GOODS-South field (HST Program ID 11359),
using 3 filters per visit ($Y_{098}$, $J_{125}$, $H_{160}$), and 2 orbits
per filter (for a total of 4800-5400s per pointing and filter).
The total area covered by the GOODS-ERS is $\sim 40$ sq. arcmin.
down to Y=27.3, J=27.4, and H=27.4 magnitudes at $5\sigma$ in an area of
$\sim 0.11$ $arcsec^2$ (corresponding to 2 times the FWHM of the images).

A full description of the imaging dataset for the ERS field and of the
reduction procedures adopted is given in \cite{grazian11}.

\subsubsection{GDS}

The CANDELS-Deep survey will cover 125 sq. arcmin. to $\sim 10$orbit
depth on the central region of the GOODS-South and Goods-North fields
at completion (\cite{koekemoer11}). At this stage, only 6 WFC3
pointings ($\sim 27$ sq. arcmin.) in the GOODS-South central region
are available, with a 3 orbit depth in the $Y_{105}$ band and $\sim
2.5$ orbits both in $J_{125}$ and in $H_{160}$ bands. In the
following, we will refer this field to Goods Deep Survey (GDS). We
added these images to our database since they reach a slightly deeper
magnitude limit ($J=27.8$ at $5\sigma$ in a circular aperture of $\sim 0.11$
$arcsec^2$) than the ERS in a comparable area, starting to
investigating the region of the parameter space dealing with faint and
extended objects: since they are faint, it is not possible to detect
them on wide surveys, i.e. the UDS, but they are very rare in
ultra-deep pencil beam surveys (i.e. HUDF).

When completed, the GDS survey will open new frontiers on the size-luminosity
studies of high redshift galaxies, thanks to its unique combination of
both area (125 sq. arcmin.) and depth ($J=28$) in the NIR wavelength range.

\subsubsection{HUDF}

The first year HUDF09 WFC3/IR dataset (HST Program ID 11563) is a
total of 60 HST orbits, observed in September 2009, in a single
pointing (\cite{oesch09,bouwens10}) in three broad-band filters (16
orbits in $Y_{105}$, 16 in $J_{125}$, and 28 in $H_{160}$). It is one
of the deepest NIR images ever taken, reaching Y=29.3, J=29.2, and
H=29.2 total magnitudes for point like sources at 5 sigma (this S/N
ratio is computed in an aperture of $\sim 0.11$ $arcsec^2$,
corresponding to two times the FWHM of the NIR images). The area
covered by the WFC3-HUDF imaging is 4.7 sq. arcmin., and the IR data
have been drizzled to the ACS-HUDF data (\cite{udf}), with a resulting
pixel scale of 0.03 arcsec and a FWHM of 0.18 arcsec.

As for the ERS, a full description of this dataset and of the
reduction procedures adopted are described in \cite{grazian11}.

\subsubsection{P12HUDF}

The HUDF09 program consists of deep observations in three different
fields (HUDF, P12, P34) on the GOODS-South region. They have been
already discussed in \cite{bouwens10c}. Here we use only data
from the P12 (hereafter P12HUDF) region to double the number
statistics at the faint end of the size-luminosity relation at $z\sim
7$. The P12HUDF field has been observed in the $V_{606}$, $I_{775}$,
and $Z_{850}$ bands by ACS (\cite{hudf05}) and in the $Y_{105}$,
$J_{125}$, and $H_{160}$ bands by WFC3 (\cite{bouwens10c}) down to a
magnitude limit of $\sim 29.1$ in the VIZ filters and $\sim 29.2$ for
the YJH ones. These data have been reduced using the same approach
described above for the GDS and HUDF fields.

\subsection{Photometry and Size of faint galaxies}

The photometric catalogs of the UDS, EGS, ERS, GDS, P12HUDF, and HUDF
fields have been derived in a consistent way. Galaxies have been
detected in the $J_{125}$ band (corresponding to rest frame 1500 \AA~
at z=7), and their total magnitudes have been computed using the
MAG\_BEST of SExtractor (\cite{sex}), using DETECT\_MINAREA=9 pixels
and DETECT\_THRESH=ANALYSIS\_THRESH=0.7. Colors in BVIZY and H bands have
been measured running SExtractor in dual image mode, using isophotal
magnitudes (MAG\_ISO) for all the galaxies. Since the FWHM of the ACS images
($0.12 \arcsec$) is slightly smaller than the WFC3 ones, we smoothed
the ACS bands with an appropriate kernel to reproduce the resolution
of the NIR WFC3 images. This ensures both precise colors for extended
objects and unbiased photometry for faint sources. Further details
on the photometric measurement can be found in \cite{grazian11}.

We used the same SExtractor catalog to measure the angular sizes of
galaxies in the samples. For this purpose we have used the half light
radius measured by SExtractor in the J band. This quantity is
the zero--order estimator of the overall shape of a galaxy that one
can adopt, and is clearly inadequate to fully describe the whole
complexity of galaxies. However, most of the $z\simeq 7$ galaxies are
very faint, close to the detection limit of our
images: in this case, the more sophisticated estimators that can be
adopted to measure galaxy morphology (e.g. the CAS system, \cite{cas}, or
the Gini coefficients, \cite{gini}) cannot be adopted, for lack of adequate
S/N.

For this reason, we adopt here the half light radius and, hereafter, we
will refer to it when generically speaking of galaxy sizes. We are
aware that, even adopting this simple estimator, this approach is
still prone to systematic effects. These will be addressed with 
detailed simulations, that are described in Sect. 4.


\section{Selecting galaxies at z=7: color criteria}

The selection of galaxies at $z\sim 7$ uses the well known ``drop-out''
or ``Lyman-break'' technique. At $6.5<z<7.5$, this feature is sampled
by the large $Z-Y$ color, as shown by a number of works using ground based
imaging (\cite{ouchi,castellano09,castellano10}) and from space
(\cite{bouwens06,Mannucci2007,hudf09,oesch09,bunker09,mclure10,bouwens10c}).
The spectroscopic confirmations of these candidates (shown in
\cite{fontana10,vanzella11,pentericci11,stark,ono}) ensures that this technique
is robust and the fraction of expected interlopers is quite low (less than
20\% at $z\sim 6$, as shown in \cite{pentericci11}).

As described in \cite{grogin11} and \cite{koekemoer11}, for the
CANDELS data on the GOODS deep survey (GDS), the $Y_{105}$, $J_{125}$,
and $H_{160}$ bands are available, as well as in the HUDF and P12HUDF
fields, while in the wide areas (UDS and EGS) only the $J_{125}$ and
$H_{160}$ bands from WFC3 are available. On the other hand, in the ERS
field the $Y_{098}$ filter is available instead of the $Y_{105}$
one. This turns out in a slightly different selection criteria for
these fields.

For the HUDF, P12HUDF, and GDS fields we thus adopted the color criteria
discussed in \cite{grazian11}:
\begin{eqnarray*}
z-Y_{105}&>&0.8,\\
z-Y_{105}&>&0.9+0.75(Y_{105}-J_{125}),\\
z-Y_{105}&>&-1.1+4.0(Y_{105}-J_{125})
\end{eqnarray*}
while for the ERS we adopted the following color criteria:
\begin{eqnarray*}
z-Y_{098}&>&1.1,\\
z-Y_{098}&>&0.55+1.25(Y_{098}-J_{125}),\\
z-Y_{098}&>&-0.5+2.0(Y_{098}-J_{125})
\end{eqnarray*}
to take into account the differences in the transmission of $Y_{098}$
and $Y_{105}$ filters.

For the non-detection in bands bluer than
$Z$, we adopt the same criteria used in \cite{castellano09,castellano10}
and in \cite{grazian11} ($S/N<2$ in all BVI
bands and $S/N<1$ in at least two of them).

Following the above criteria, we select 20 candidates
in the HUDF field down to $J<29.2$, while in the ERS we find 22 z-dropout
candidates at $J<27.4$, respectively. Their characteristics are described in
Table 1 and 2 of \cite{grazian11}. For the GDS and P12HUDF we recover
21 and 23 galaxies down to $J<27.8$ and $J<29.2$, respectively.

For the UDS and EGS fields, where the only photometry available
from space is in $V_{606}$, $I_{814}$, $J_{125}$, and $H_{160}$ bands,
we adopt the $I_{814}$-dropout color
selection, which gives a more extended redshift window for selecting galaxy
candidates ($6.4<z<8.5$). In particular, the color criteria adopted are:
\begin{eqnarray*}
I_{814}-J_{125}&>&2.0,\\
I_{814}-J_{125}&>&1.4+2.5(J_{125}-H_{160})
\end{eqnarray*}
with $J_{125}\le 26.7$ and non detected in the $V_{606}$ band
($S/N(V_{606})<1.5$),
which is satisfied by 46 galaxy candidates in the UDS and 21 in the
EGS (the present observation of this field covers only half of the expected
FoV).

In appendix A we provide the tables with position, $J_{125}$ band magnitude
and size of the $z\sim 7$ candidates found in the GDS, P12HUDF, EGS,
and UDS fields.

\begin{figure}
\centering
\includegraphics[width=8cm]{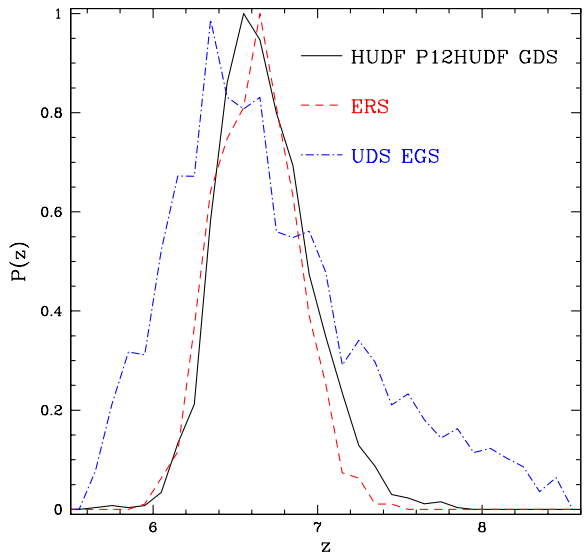}
\caption{The selection function in redshift for the different color
criteria adopted in the six CANDELS and HUDF fields.
The peaks of the distributions have been normalized to 1 to ease the
comparison.
The selection function for the CANDELS wide surveys (UDS and EGS) is
more extended due to the limited set of filters used (VIJH) with respect
to the filters used in GOODS Deep or HUDF surveys (BVIZYJH).
}
\label{selefunctz}
\end{figure}

Fig. \ref{selefunctz} shows the selection functions in redshift for the
different color criteria described above. They have been derived using the
simulations described in the next section, and applying to the
simulated catalogs the same color selections described above for real dropouts.
The selection function for the CANDELS wide surveys (UDS and EGS) is
more extended due to the limited set of HST filters used (VIJH) with respect
to those used in GOODS Deep or HUDF surveys (BVIZYJH).
In particular, in the UDS and EGS surveys we adopt the $I_{814}$ drop-out
criterion, which selects lower-z candidates than the classical
$Z_{850}$ drop-out galaxies. Indeed, the lack of the HST Y band for UDS and
EGS results in a
redshift selection function which is more extended towards higher-z objects.
The redshift distributions expected for HUDF and for ERS are similar,
despite the different Y band filters adopted.
The mean redshifts of all the distributions are around 6.7, indicating
that there's no difference in the typical redshifts of selected dropout
galaxies in these six surveys.

\subsection{The reliability of candidates selected on the CANDELS Wide surveys}

In the CANDELS wide surveys analysed here (UDS and EGS), the reliability of
$z\sim 7$ galaxies candidates can be hampered by the limited set of
HST images (VIJH) available. We have carefully investigated the properties
of our dropout candidates in the UDS field using both
the photometric redshift technique based on SED fitting and the stacking
of HST images of all the candidates to check the expected non detection in the
optical images.

To compute the photometric redshifts for our candidates in the UDS field,
we complement the VIJH HST photometry with the ground based images
available from CFHT in the U band, from Subaru in the BVRiz filters
and from UKIRT in the JHK bands. We added also the IRAC photometry both from
SpUDS and SEDS programs, using the TFIT software to match the resolution of
HST to the PSF of ground based images or of space based images by Spitzer.
The adopted technique for the derivation of the photometric catalog is
described in detail in \cite{galametzuds}.

The photometric redshift analysis used to selected the high-redshift
candidates in the UDS field was performed using the template-fitting
code developed by \cite{mclure11}. For the purposes of this study we
employed the \cite{bc03} stellar evolution models, with
metallicities ranging from solar ($Z_{\odot}$) to $1/50$th solar
($0.02Z_{\odot}$). Models with instantaneous bursts of star-formation,
constant star-formation and star-formation rates exponentially declining
with characteristic timescales in the range 50~Myrs~$<~\tau~<$~10~Gyrs
were all considered. The ages of the stellar population models were
allowed to range from 10 Myrs to 13.7 Gyrs, but were required to be less
than the age of the Universe at each redshift. Dust reddening was
described by the \cite{Calzetti2000} attenuation law, and allowed to
vary within the range $0.0<A_{V}<2.5$ magnitudes. Inter-galactic medium
absorption short-ward of Ly$\alpha$ was described by the \cite{Madau1995}
prescription, and a Chabrier IMF was assumed in all cases.
For each model of this grid, we have
computed the expected magnitudes in our filter set, and found the
best--fitting template with a standard $\chi^2$ normalisation.

Within the UDS sample, only 27 galaxies have a robust photometric redshift
$z\ge 6.8$, while the remaining galaxies are either consistent with having
a slightly
lower value ($6.3\le z\le 6.8$), or have two comparable solutions for the
photometric redshifts (one at $z\sim$2 and the other at $z\sim 7$).
These 27 candidates are part of another work on the luminosity function of
bright z-dropout LBGs and will be described in detail in \cite{mclure12},
with a full description of the photometric redshifts used here.
We decided to keep the full sample of 46 candidates for UDS in our work,
after checking that they have comparable properties (both in luminosity and
half light radius) to the sub-sample of the 27 candidates of \cite{mclure12}.
We repeat the same check on EGS, and we find that 13 galaxies out of 21 have
$zphot\ge 6.8$.

\begin{figure}
\centering
\includegraphics[width=8cm]{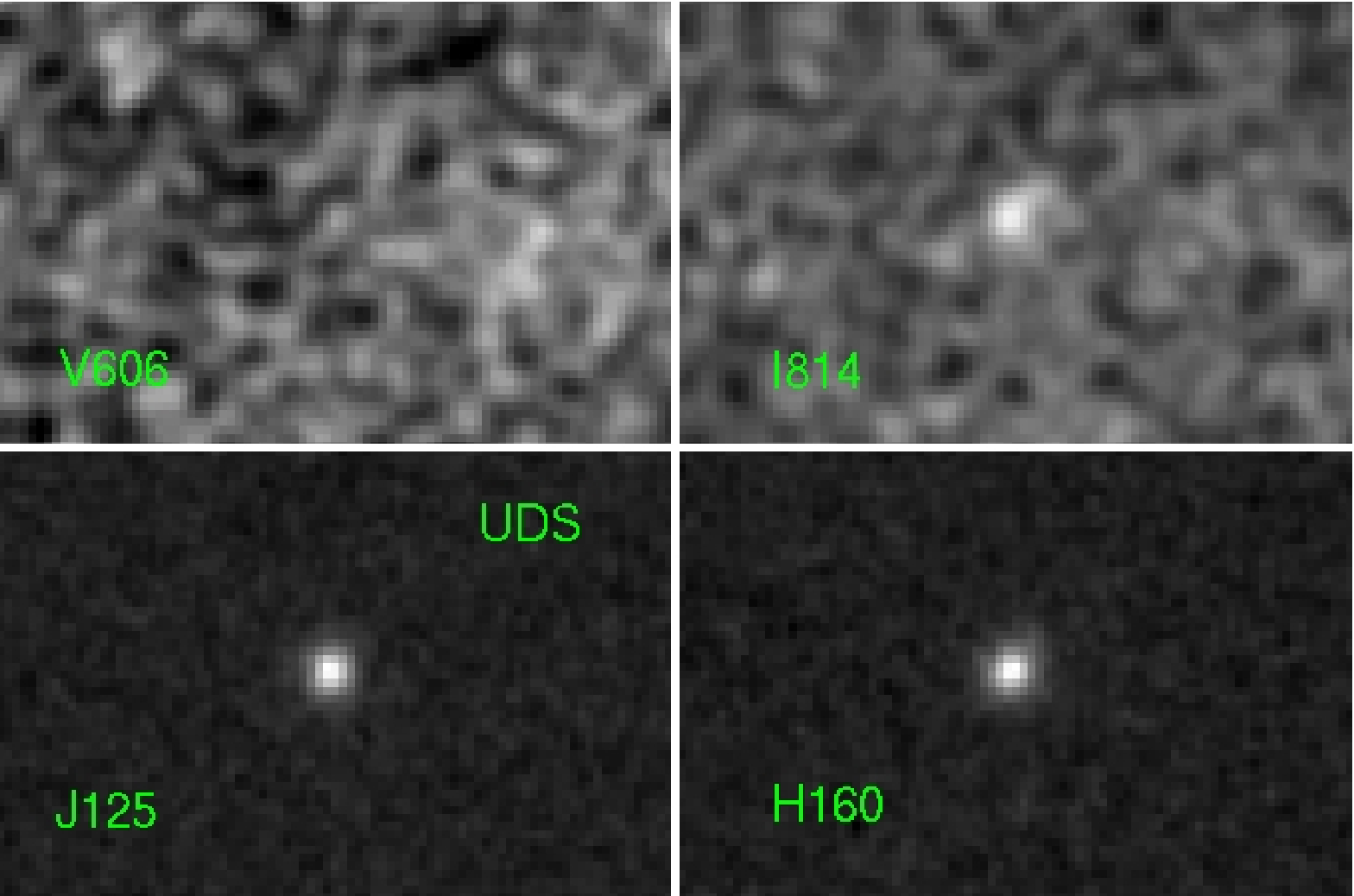}
\includegraphics[width=8cm]{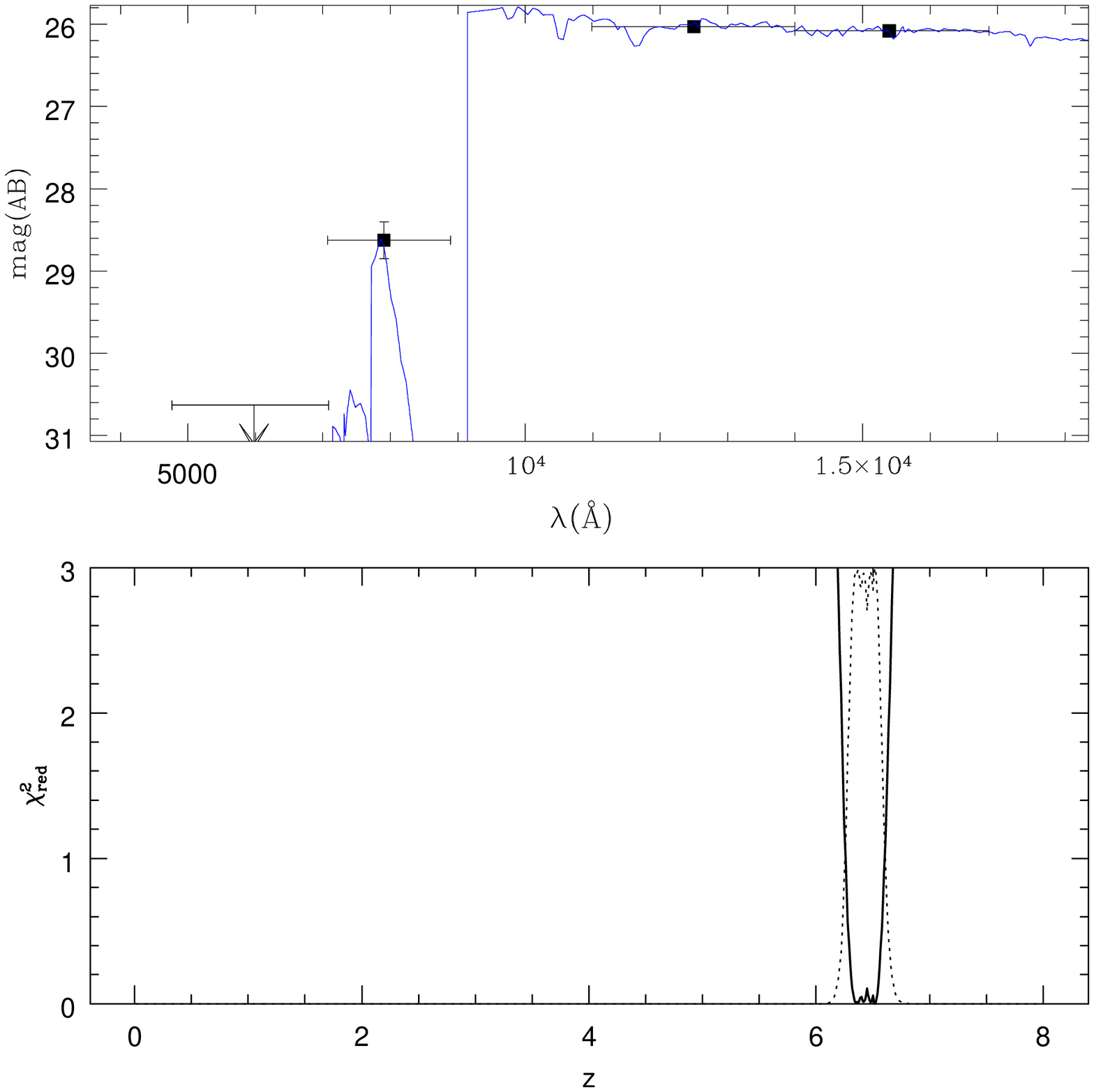}
\caption{{\it Top:} the image stacking of the 46 $I_{814}$-dropout
candidates in the UDS field for $V_{606}$, $I_{814}$, $J_{125}$, and
$H_{160}$ bands. {\it Middle:} the VIJH photometry of this image
stacking (black dots) has been fitted with an extended library of
synthetic galaxies, and the best fit (blue line) is consistent with a
photometric redshift $z\ge 6.5$, with no secondary peak at $z\sim
1-2$. The model templates are computed with the spectral synthesis
models BC03 (\cite{bc03}), and chosen to broadly
encompass the variety of star--formation histories, metallicities and
extinction of real galaxies, as described in detail in
\cite{mclure11}.  At z=6.5 the Lyman continuum absorption is at an
observed wavelength of $\lambda=6840$ \AA, and a faint emission in the
$I_{814}$ filter can be still compatible with a partially neutral IGM
at this redshift. {\it Bottom:} The continuous curve is the reduced
$\chi^2$ as a function of photometric redshift $z$, while the dotted
line is the probability P(z) rescaled to have the peak at the same
level of $\chi^2$, to improve visibility.
}
\label{UDSstack}
\end{figure}

Fig.\ref{UDSstack} shows the weighted mean of the 46 $I_{814}$-dropout
candidates in the UDS field for $V_{606}$, $I_{814}$, $J_{125}$, and
$H_{160}$ bands. The
stack image has no detection in the $V_{606}$ band, and a very faint
detection in the $I_{814}$ band. The fit to the SED of the stack indicates
that the photometric redshift is consistent with $zphot\sim 6.5-7.5$
and no secondary peak at $z\sim 1-2$ is present. Restricting the
sample to galaxies with magnitude $J_{125}\le 26.2$ and angular size
$Rh\ge 0.2$ arcsec, we obtain essentially the same stacked SED and photometric
redshift probability as that for the whole sample. This indicates that the
46 candidates in the UDS field are robust against contamination by low-z
interlopers, and
bright and extended $I_{814}$-dropouts are not dominated by foreground
sources mimicking our color criteria.
In the other fields, where the depth of HST images
are similar (EGS) or deeper (ERS, GDS, HUDF, P12HUDF), we expect as well
a reduced contamination by lower-z sources.

\section{Simulations for estimating completeness and systematic effects}

While the selection criteria described above are formally designed to
select a pure sample of high-z candidates, they are in practice
applied to very faint objects, typically close to the limiting depth
of the images. At these limits, systematics may significantly affect
their detection and the accurate estimate of their colors or
apparent dimensions. To take into account all the systematic effects
(completeness, photometric scatter, size scatter) involved in the
size-luminosity relation, we carried out a set of detailed
simulations.

These simulations have two main goals. The first is to estimate the
{\it incompleteness} that affects the detection of faint galaxies as a
function of their size. The incompleteness arises not only from the
difficulty of detecting faint sources, but also from the effect of
noise in the many bands that we need to select high-z candidates. As
expected, we become severely dominated by incompleteness as we attempt
the detection of extended sources at the faintest magnitudes, and this
effect must be taken into account when trying to infer the intrinsic
size--luminosity relation.

The second goal is to quantify the systematic biases in the measure of the
half--light radius provided by SExtractor, as a function of magnitude
and galaxy size.
These systematic biases affect the estimate of the
distribution of galaxy size in a
non-negligible way, and therefore also need to be treated carefully
in the analysis.

In this section we describe in some detail the simulations adopted to
estimate these systematic effects and the technique that we adopt to
include them in the derivation of the correct size distribution. The reader
directly interested in the observational results may skip this section
and proceed to Sect. 5.

\subsection{Completeness}

We follow the procedure described in \cite{grazian11} and in
\cite{castellano09,castellano10} to obtain a realistic catalog of simulated
high redshift sources. We first produce a set of UV absolute magnitudes
and redshifts $(M_{1500},z)$ according to an evolving Luminosity
Function by \cite{castellano10}, and then convert it into a set of
predicted magnitudes using the BC03
models with a range of ages, metallicities, dust content and
Ly$\alpha$ emission as in in \cite{grazian11}. We have also added the
intergalactic (IGM) absorption using the average evolution as in \cite{fan06}.
The redshift range adopted for the simulated galaxies in all the fields is
$5.5<z<8.5$ while the absolute magnitudes $M_{1500}$ run from -17 to -23.
The input distribution for the axial
ratio parameter $b/a$ is assumed flat from 0 to 1.

For each simulated galaxy, an input half-light radius has been assigned
by selecting at random from a uniform distribution between 0.0 and
1.0 arcsec. The two-dimensional profiles adopted during the
simulations are typical of disk galaxies, i.e. an exponentials.
In Appendix B we show the results of our simulations under the
assumption of a Sersic profile of index $n=4$.
Previous works (e.g. \cite{ferguson04}) dealing with the
size-luminosity relation at lower-z, assumed a mix of morphological
models, with a fraction of 70\% spiral and 30\% ellipticals at
z=3-4. From the observational point of view, \cite{rav06} find that
40\% of the LBGs at $2.5<z<5$ have light profiles close to
exponential, as seen for disk galaxies, and only $\sim$30\% have high
Sersic index, as seen in nearby spheroids. They also find a
significant fraction ($\sim$30\%) of galaxies with multiple cores or
disturbed morphologies, suggestive of close pairs or on-going galaxy
mergers. Using the ultradeep images of the HUDF in BVIZ bands by ACS,
\cite{hathi08b} found that the sum of the images of all LBGs selected
at $4\le z\le 6$ are well fit by Sersic profiles with an index
$n<2$, indicating that these galaxies follow a disk-like profile in
their central region, as recently confirmed by \cite{fathi12} (but see
\cite{gg03} for a different interpretation).
Moreover, recent results from spectroscopically
confirmed LBGs at $z\le 3$ (\cite{nagy,law}) and $z\sim 5$
(\cite{douglas10}) show that the light profile of these galaxies are
not represented by an elliptical morphology, and the brightest
galaxies are typically described by an exponential disk. Based on
these considerations, from now on we {\it assume} that galaxies at
$z\sim 7$ are typically approximated by an exponential disk morphology
and that there is no galaxy with an $r^{1/4}$ profile at such high-z.
Though this is only a rough approximation, there are indications, at
redshifts lower than 7, that it is a realistic assumption.

The synthetic galaxies are placed at random positions in the
real 2-dimensional FITS images, avoiding positions where 
where real galaxies or stars are observed. To this aim, the
segmentation image created by SExtractor during the detection of real
objects in the observed J band has been used for each field. To avoid
an excessive and unphysical crowding in the simulated images, we have
included only 200 objects of the same flux and morphology each time. We
then perform the detection in the synthetic images using SExtractor
with the same parameters adopted for the real images. We simulate all
the available bands, i.e. from the B to the H bands for the ERS and
HUDF fields while for the UDS and EGS fields we have simulated only the VIJH
filters from HST observations. We repeated the simulation until a
total of at least $5\times10^5$ objects were tested for each of the
six fields described above.

\begin{figure}
\centering
\includegraphics[width=8cm]{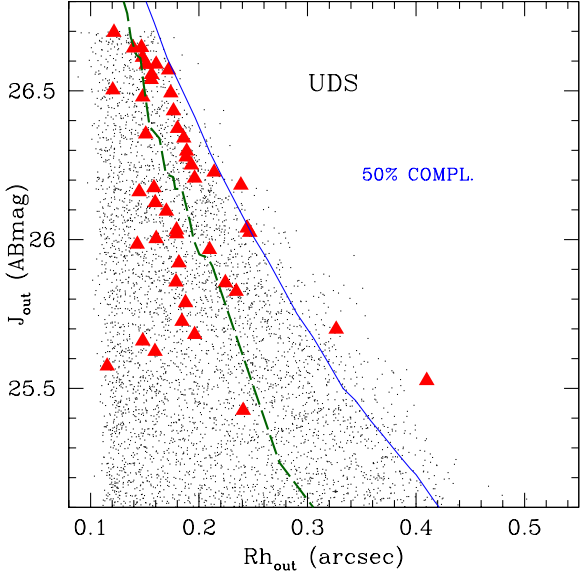}
\caption{The observed J magnitude vs size of simulated (small black dots) and
observed galaxies (red triangles) at $z\sim 7$ for the UDS field.
The solid blue line shows the 50\% completeness level
for an input simulated profile of disk galaxy, while the dashed green
line is the same limit for a Sersic profile with n=4.}
\label{UDSmag_rh}
\end{figure}

These simulations first provide the detection completeness as a
function of the input half light radius and input total magnitudes.
The typical output is shown in Fig.\ref{UDSmag_rh}, where we plot the
measured total magnitude $J_{out}$ and half-light radius $Rh_{out}$ in
the case of the UDS field. Small dots show the position of the
simulated objects, while big red triangles are the observed $z\sim 7$
LBG candidates. Since the simulated input catalog contains objects of
all sizes up to 1'' in equal proportions, 
Fig.\ref{UDSmag_rh} clearly shows that, at a given
magnitude, the fraction of detected objects drops above some critical
size. We use these simulations to compute the 50\% completeness
threshold as a function of magnitude, which is shown as a blue solid
line in Fig.\ref{UDSmag_rh}.

We repeated the same analysis for all the six fields in our
survey and we show the output in Fig.\ref{FAINTmag_rh}
of Appendix B, for the EGS, ERS,
GDS, P12HUDF and HUDF fields, respectively.
Results are qualitatively very similar, the only major
difference being that the 50\% completeness curve shifts to fainter
magnitudes on the deeper fields, as expected.

We also explored the effects of our assumption for the galaxy
surface-brightness profiles. We repeated the analysis adopting as input for our
simulations a Sersic profile with index $n=4$, and show the
completeness obtained with this profile in Fig.\ref{UDSmag_rh}
(dashed line), comparing it with the previous result for an exponential
profile (solid line). As expected, the 50\% completeness threshold
occurs at lower radii (for a given magnitude) with the elliptical
profile than with the exponential one. It is worth noting that the
elliptical profile cannot explain the presence of galaxies brighter
than $J=26$ and larger than 0.3 arcsec in half light radius that we
are finding in the UDS field. It is useful to stress that the completeness
due to a Sersic profile with index $n=4$ carried out here is only a simple
exercise and in the following sections of this paper we make the reasonable
{\it assumption} that all the $z\sim 7$ candidates found in the CANDELS
and HUDF fields have an exponential disk profile.

\begin{figure}
\centering
\includegraphics[width=8cm]{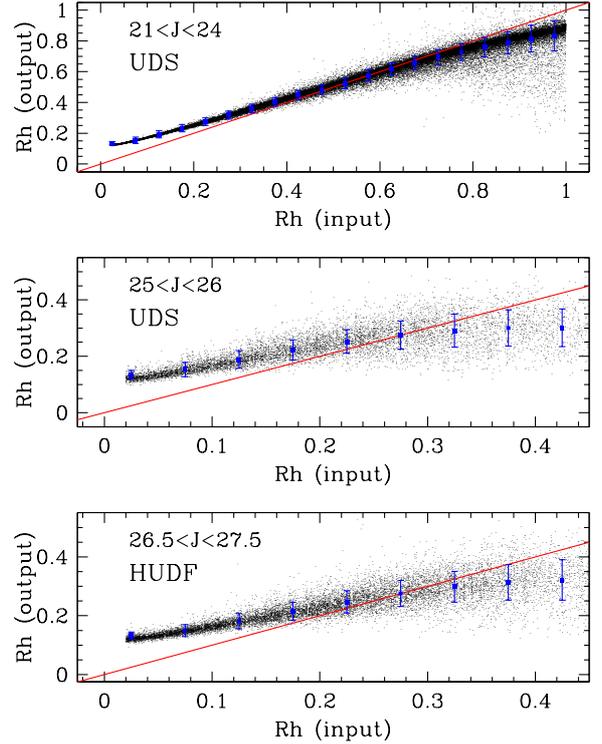}
\caption{The results of our simulations for $z\sim 7$ galaxies with exponential
disk profiles in the UDS and HUDF fields.
The plot shows the comparison between the half--light radius (in arcsec) as
measured by SExtractor as a function of the input one. The upper and
lower panel refer to different total magnitudes in two different
simulations, as reported in the legend. The red line shows the
identity relation. The blue points and errorbars show the average
value and the relevant r.m.s. of the output half--light radius.
At small sizes the output half light radius is typically larger than the
input one due to the convolution with the instrumental PSF carried out
during the simulations.
Note that the limits of the top panel are different from the ones in the
middle and bottom panels.}
\label{rhl}
\end{figure}

A second output of our simulations is the estimate of the biases in the
measurement of the half--light radius as a function of luminosity and size.
We show the results in Fig.\ref{rhl}, where we plot the measured
$Rh$ as a function of the input one, in three magnitude ranges. The
upper and middle plots refer to the UDS field, while the bottom plot describes
the same simulation for the HUDF field.
The upper panel shows the result for
relatively bright objects ($21\le J\le 24$).
While no $z\simeq 7$ galaxies are found in
this magnitude range, it is instructive to see that systematic effects
are important even at the bright end.
As expected, the measured size of the objects cannot be
smaller than the instrumental PSF (which is about 0.18'' of FWHM in the J
band, corresponding to an half light radius of 0.11'').
Because of the convolution with the PSF, all objects
intrinsically smaller than $\simeq 0.2$'' are biased high by the
SExtractor estimate. Above 0.4'', the opposite starts to occur. This
is due to the detection algorithm adopted. In SExtractor an object is detected
only if a given number of pixels have an intensity above a defined threshold,
and the half light radius is computed using only those pixels.
For faint and quite extended galaxies, the surface brightness far from the
center is very low and the galaxy merges into the noise, with the effect of
underestimating both the total flux and the effective size for these
objects. The middle and bottom panels of Fig.\ref{rhl} show the same
simulations carried out in two magnitude ranges that are critical for the
aims of this work (note that the half light radius range in the x and y-axis
is different from the one in the top panel).

\begin{figure}
\centering
\includegraphics[width=8cm]{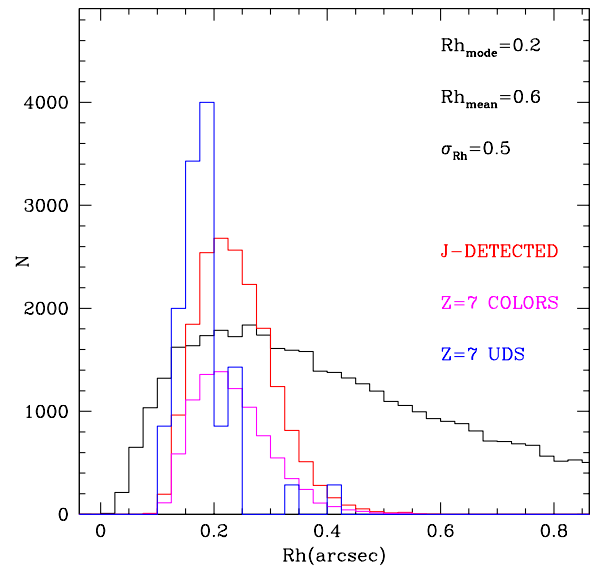}
\caption{The size distributions of simulated and
observed galaxies at $z\sim 7$ for the UDS field. The black histogram
represents the input log-normal function with mean half
light radius of 0.6 arcsec and a $\sigma_{Rh}$ of 0.5, resulting in a
peak at 0.2 arcsec (mode). This is not the best fit for the observed size
distribution, but only an example to show the effects of incompleteness
at large half light radii.
The input galaxies are simulated down to
a magnitude of $J=26.7$, which is the nominal limit for the UDS field.
The red histogram is the size distribution
of all galaxies detected in the J band UDS image, irrespectively of
magnitudes and colors, while the magenta histogram shows only the
galaxies at $z\sim 7$ selected by our color criteria. Last, the blue
distribution represents the observed galaxy sizes in the UDS, scaled
in normalization to match the number of simulated galaxies.}
\label{simulstep}
\end{figure}

The combined effect of the two systematic biases described above (on
the detection and on the estimated size) has a measurable impact
on the observed distribution of galaxy sizes. An example is given in
Fig.\ref{simulstep}. We assume that the real input distribution is a
log-normal with mean half light radius of 0.6 arcsec and a
$\sigma_{Rh}$ of 0.5 (black histogram) which has a mode (peak) at
0.2 arcsec. These values are not representative of the best fitting
parameters for the UDS half light radius distribution, but they are randomly
chosen just to provide an example with an extended tail, to show the
selection effects at large $Rh$.

The distribution of the measured sizes (red histogram) of the
simulated objects deviates from the input distribution in two respects. 
At small sizes, the
distribution is truncated below 0.1-0.15'' and the peak is moved to a
slightly larger value, because of the convolution with the PSF. At
large sizes, a clear cut above half light radius of 0.4 arcsec is
evident, due to the detection incompleteness discussed in
Fig.\ref{UDSmag_rh}.
We also note that when the color criteria for $z\sim 7$ are applied,
the amplitude of the histogram is reduced but its shape is not changed
(magenta distribution), implying that the color selection is not
significantly affecting the half--light distribution of the simulated objects.
The $Rh$ distribution for real objects, scaled
in normalization to match the number of simulated galaxies,
is shown by the blue histogram in Fig.\ref{simulstep}.

\subsection{Finding the best fit to the observed distribution of
galaxy half--light radius}

We finally use our simulations to recover the true size distribution
of z=7 galaxies, under the assumption of a particular functional form.
We parameterized this as a log-normal function in half light radius, with the
two parameters $Rh_{mode}$ and $\sigma_{Rh}$ that are independent from
the input luminosities for each simulated sample. This
assumption is based on the fact that the observed size distribution at
different redshifts is characterized by a well-defined peak and a tail
towards more extended objects (e.g. \cite{nagy}). The log-normal
functional form naturally fits this shape, and it is characterized by
two parameters, $Rh_{mode}$ and $\sigma_{Rh}$, the peak and the
dispersion of the log-normal half light radius distribution.
Moreover, from the theoretical point of view, the log-normal function
is expected as a natural distribution for galaxy sizes (\cite{mmw98}).

In order to recover the intrinsic shape of the distribution we have
adopted a Maximum Likelihood (ML) approach. For any choice of the free
parameters of the log-normal function ($Rh_{mode}$ and $\sigma_{Rh}$),
the resulting intrinsic distribution
is first convolved with the observational biases (as described above,
see Fig.\ref{simulstep}) to get the expected number of sources as a
function of the measured half light radius. The total number of observed
galaxies has been matched to the simulated one. Then, a Poisson likelihood
is computed
comparing the simulated and observed $Rh$ distributions, using
the usual formula:
\begin{equation}
\label{eq:ml}
{\cal L} = \prod_{i} e^{-N_{exp,i}} \frac{(N_{exp,i})^{N_{obs,i}}}{(N_{obs,i})!
}
\end{equation}
where $N_{obs,i}$ is the observed number of sources in the half light
radius interval $i$, $N_{exp,i}$ is the expected number of simulated sources in
the same half light radius interval, and $\Pi_{i}$ is the product
symbol.

Because the observed J-band magnitude range of our $z\sim 7$ candidates
in the six CANDELS fields is roughly two magnitudes, and even
larger for HUDF and P12HUDF, we limit both the observed and the
simulated samples to a relatively small magnitude interval for the
maximum likelihood computation. For each field, we use only galaxies
(both observed and simulated) in the magnitude range $25.6\le J\le
26.6$ to compute the likelihood for bright galaxies; for the
intermediate magnitude bin we limit the interval to $26.6\le J\le
27.6$, while for the faintest bin we adopt $27.6\le J\le 28.6$ as
limits.  Even though a size-luminosity relation holds at $z\sim 7$, as
we will show later in this paper, this choice ensures that the
change in $Rh_{mode}$ inside each analysed magnitude interval
is small.

\begin{figure}
\centering
\includegraphics[width=8cm]{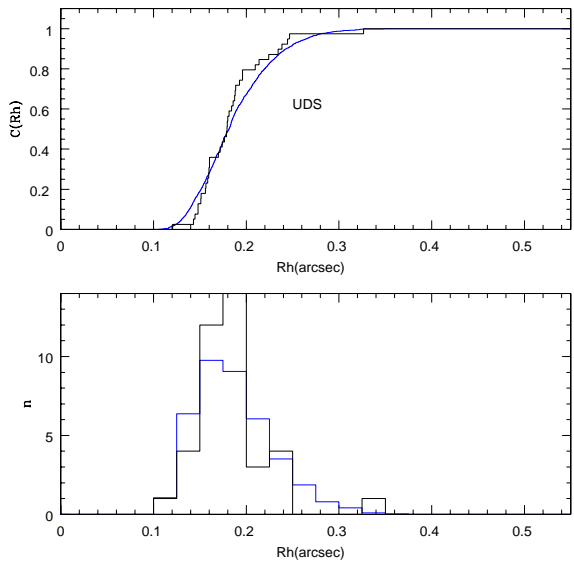}
\caption{The best fit for the size distribution of the
observed galaxies at $z\sim 7$ in the
magnitude range $25.6\le J\le 26.6$ for the UDS field.
{\it Top:} the cumulative distribution for the observed galaxy sizes
(black histogram)
is compared with the same for the simulated galaxies (blue curve), and a K-S
p-value is computed.
{\it Bottom:} The observed (black) and simulated histograms (blue),
as described in the top panel, are compared with a Maximum Likelihood approach.
}
\label{bestfitUDS}
\end{figure}

For each combination of the two parameters $Rh_{mode}$ and
$\sigma_{Rh}$, the peak and the dispersion of the log-normal half
light radius distribution, we compute the maximum likelihood as
described in Eq.\ref{eq:ml}.
A typical example is provided in Fig.\ref{bestfitUDS}, that
shows the best fit distribution resulting from comparing observations
and simulations in the UDS field, using $z\sim 7$ galaxies in the
magnitude range $25.6\le J\le 26.6$.
We plot both the differential (bottom)
as well as the cumulative size distributions (top). This figure shows the
observed distributions of half light radii (black segmented
line) and a comparison is made with the cumulative distributions
resulting from the simulations discussed above (blue curve) after all
the systematic effects are taken into account, especially the convolution of
the intrinsic galaxy shape with the observed PSF.
We compute the Kolmogorov-Smirnov p-value
by comparing the observed and the simulated cumulatives.
Since the K-S test is not sensitive to the details of the distribution,
we prefer
the Maximum Likelihood approach (bottom panel), as we discussed above,
to find the best agreement between simulations and observations.

For each of the six fields we scan a grid in two parameters,
$Rh_{mode}$ and $\sigma_{Rh}$. The grid extends from 0.01 to 0.4
arcsec in $Rh_{mode}$ (corresponding to an interval of 0.02-1.0 arcsec
in $Rh_{mean}$) and from 0.02 to 1.0 arcsec in $\sigma_{Rh}$, for a
bin size of 0.01 arcsec in $Rh_{mode}$ and 0.02 in $\sigma_{Rh}$.  The
choice of parameters ($Rh_{mode}$, $\sigma_{Rh}$) that maximizes the
likelihood is considered as the best fit for the real input
distribution. To improve the signal-to-noise ratio of the fitting
procedure, we combine the likelihoods of the fields within the same
J-band magnitude range. In addition to the best fit values, the ML
approach allows us to define the allowed confidence intervals on the free
parameters $Rh_{mode}$ and $\sigma_{Rh}$.

\begin{figure}
\centering
\includegraphics[width=8cm]{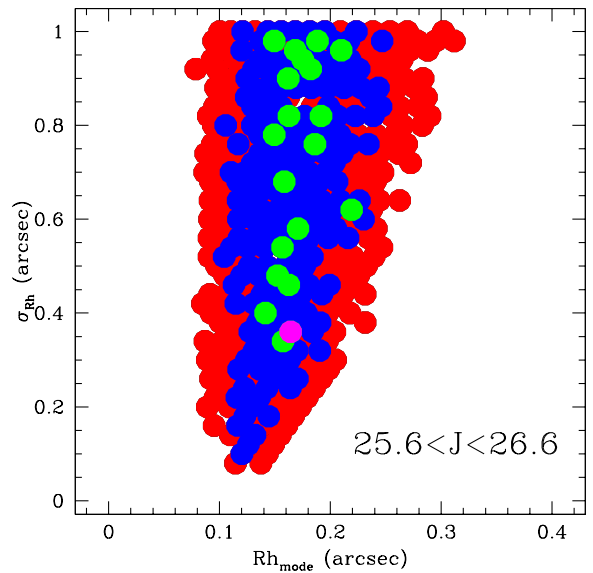}
\caption{The confidence levels obtained with the maximum likelihood approach
for the size distribution on the magnitude interval $25.6\le J\le 26.6$.
The best fit is indicated by the magenta point
while the green, blue, and red regions define the uncertainties at 68\%, 95\%,
and 99.7\% (1,2, and 3 sigma) confidence level, respectively.
The $\sigma_{Rh}$ is basically unconstrained, while the parameter
$Rh_{mode}$ is well constrained between 0.14 and 0.22 arcsec at 1 sigma.}
\label{regionBRIGHT}
\end{figure}

Fig.\ref{regionBRIGHT} shows the likelihood distribution for the
combined UDS, EGS, ERS and GDS fields for galaxies in the magnitude
range $25.6\le J\le 26.6$. The best fit
is indicated by the magenta point while the green, blue, and red
regions define the uncertainties at 68\%, 95\%, and 99.7\% (1,2, and 3
sigma) confidence level, respectively.
In Fig.\ref{regionBRIGHT} $Rh_{mode}$ and $\sigma_{Rh}$ represent
the intrinsic parameters of the input log-normal distribution before
convolving it with all the observational effects (PSF convolution,
noise, detection and size measurements).
Fig.\ref{regionallfaint} in Appendix C shows
the same plot for the other J band intervals, namely the combination of
all the fields for the intermediate magnitude bin $26.6\le J\le 27.6$,
and the combined GDS, P12HUDF and HUDF fields for the
faintest bin $27.6\le J\le 28.6$, respectively.
From this statistical analysis we derive the best fit values $Rh_{mode}$ and
$\sigma_{Rh}$ used in the following, together with their 68\% confidence level
intervals.

In Fig.\ref{regionBRIGHT} the peak of the half light distribution
($Rh_{mode}$) for the bright sample ($25.6\le J\le 26.6$) is well
constrained by the present observations, while it is not possible to
put reasonable limits for the spread of the size distribution
($\sigma_{Rh}$). The reason for this behavior is clear going back to
Fig.\ref{simulstep}: input distributions with extended tails like the
black histogram, after the convolution with simulated effects, are
affected by a strong incompleteness at large sizes (red and magenta
histograms) and thus the present data in the UDS and EGS fields, even
after detailed comparison
with simulations, cannot provide stringent constraints to the
extensions of the intrinsic distributions in size.


\section{Results}

\begin{figure}
\centering
\includegraphics[width=9cm]{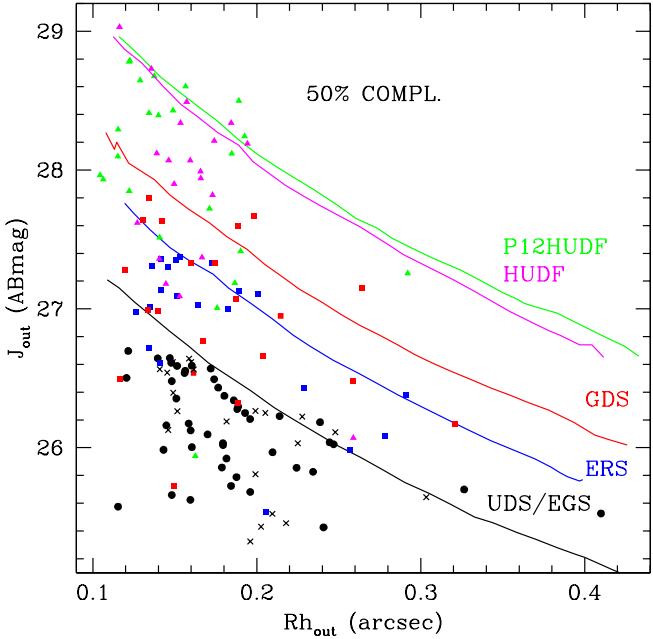}
\caption{The observed J magnitude vs size of
candidate galaxies at $z\sim 7$ for all the six fields analysed.
The black circles indicates galaxies in the UDS field, while crosses
are associated to objects in the EGS field.
The solid lines show the 50\% completeness levels of the six different surveys
for an input simulated profile of exponential disk galaxy.
The galaxy size ($Rh_{out}$) is the observed half light radius in arcsec
measured by SExtractor and it is not deconvolved by the PSF.
}
\label{ALLmag_rh}
\end{figure}

\subsection{Faint galaxies are always small}

First of all, we present the relation between galaxy half light radii
and fluxes for our $z\sim 7$ sample. Fig.\ref{ALLmag_rh}
summarizes the result for the observed size-magnitude distribution at
$z\sim 7$ for all the six fields investigated in this work. The solid
lines show the 50\% completeness levels of the six different surveys
for an input simulated profile of exponential disk galaxy, as
described above. The galaxy size ($Rh_{out}$) is the observed half
light radius in arcsec measured by SExtractor, without any attempt to
deconvolve it for the effect of the PSF.

It is interesting to note in this plot the lack of galaxies in the
region of large sizes but faint magnitudes (or low surface brightness)
objects, namely fainter than J=26.6 and larger than 0.2 arcsec
(corresponding to a physical dimension of 1.15 kpc at z=7). Two
notable exceptions are 2 galaxy candidates, one in the GDS and the
other in the P12HUDF field. Visual inspection of these two objects
indicates that they are both close to bright and extended galaxies,
thus their magnitudes or sizes could be affected by the presence of
the bright interlopers. They do not show any evidence of merging or
clumpy morphology. We included these two galaxies in the Maximum
Likelihood analysis described above. Excluding these two objects from
Fig.\ref{ALLmag_rh}, we detect a clear lack of faint and extended
galaxies at $z\sim 7$, corroborated by very large statistics (153
candidates over 6 fields) and down to very faint magnitude limits
($J\sim 29$). Limiting our analysis to $26.6<J<27.6$, where the lack
of extended galaxies is evident and where we are reasonably complete
with our survey, we have 41 galaxies.
Moreover, as shown in Fig.\ref{rhl} (bottom
panel), in this magnitude range the true value of the galaxy size is
not particular underestimated, and from our simulations we expect to
find exponential disk galaxies with $Rh=0.3-0.4$ arcsec, if they
exist. None of them have been found on HUDF and P12HUDF.

From the best fit results of Fig.\ref{regionBRIGHT} and
Fig.\ref{regionallfaint} in Appendix C it is clear that the parameter
$\sigma_{Rh}$, which regulates the amplitude of the log-normal
distribution in size, is basically unconstrained in almost all the fields
(see also Table \ref{tab:resu}).
It is also worth noting that in Fig.\ref{regionallfaint} we have a marginal
indication (68\% confidence level) that the parameter $\sigma_{Rh}$
is less than 0.14 arcsec on the magnitude interval $26.6\le J\le 27.6$.
If confirmed, this could strengthen the evidence of a lack of extended
galaxies at z=7 in the faint luminosity regime ($J\sim 27-28$).
Conversely, as shown in Fig.\ref{ALLmag_rh},
at magnitude brighter than $J=26.6$ (i.e. $M_{1500}\sim -20.5$
for z=7) galaxies can be as extended as $Rh=0.4$ arcsec, or 2.3 kpc
physical, while fainter than this limit the sizes are always less than
0.2 arcsec.

\subsection{The best fit log-normal distribution}

\begin{figure}
\centering
\includegraphics[width=8cm]{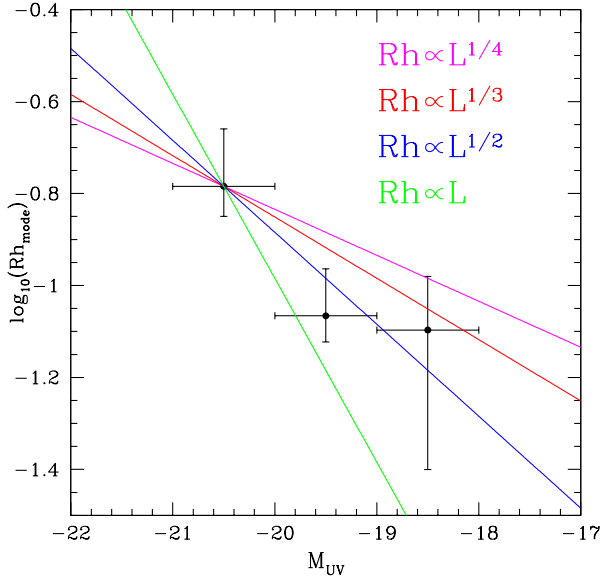}
\caption{The trend of galaxy size (in arcsec and logarithmic scale)
with UV absolute magnitude (1500\AA rest frame) at z=7 and the comparison
with some simple power-law relations. The data point at $M_{UV}\sim -20.5$ is
related to galaxies in the magnitude range
$25.6\le J\le 26.6$, the one at -19.5 is derived
using objects with $26.6\le J\le 27.6$, while the faint bin at
-18.5 is the combination of GDS, HUDF, and P12HUDF fields (where the maximum
likelihood has been computed restricting to $27.6\le J\le 28.6$).}
\label{sizelum}
\end{figure}

The visual impression derived from Fig.\ref{ALLmag_rh} is confirmed by
the robust analysis with the ML method. Adopting the formalism described
above, we fit the observed size distributions in the six fields with a
log--normal function, taking into account all the observational
biases. Results are summarised in Fig.\ref{sizelum}, where we plot
the peak (mode) half light radius (of the intrinsic distribution) at
different luminosities of the $z\sim 7$ galaxy
candidates in each survey. The brightest point refers to galaxies in the
magnitude range $25.6\le J\le 26.6$,
while the middle point at $M_{UV}\sim -19.5$ is the result of the
joint Maximum Likelihood using objects with
$26.6\le J\le 27.6$. In the faint bin we combine
the Likelihood values for the GDS, HUDF, and P12HUDF fields
at $27.6\le J\le 28.6$. The vertical error bars are the 1$\sigma$
uncertainties on $Rh_{mode}$ found with the simulations described
above, while the horizontal error bars show the minimum-maximum range in
luminosity of the sample. The results for each magnitude interval
are summarized in Table \ref{tab:resu}.

\begin{table*}
\caption{Size-Luminosity Results at $z\sim 7$}
\label{tab:resu}
\centering
\begin{tabular}{l c c c c c c c c c c}     
\hline\hline
Interval & $N_{cand}$ & $M_{UV}$ & $M_{UV}^{min}$ & $M_{UV}^{max}$ &
$Rh_{mode}$ & $Rh_{mode}^{low}$ & $Rh_{mode}^{up}$ & $\sigma_{Rh}$ &
$\sigma_{Rh}^{low}$ & $\sigma_{Rh}^{up}$\\
\hline
 & & & & & (arcsec) & $-1\sigma$ & $+1\sigma$ & (arcsec) & $-1\sigma$ & $+1\sigma$\\
\hline
$25.6\le J\le 26.6$ & 63 & -20.5 & -21.0 & -20.0 & 0.16 & 0.14 & 0.22 & 0.36 & 0.34 & 0.98\\
$26.6\le J\le 27.6$ & 36 & -19.5 & -20.0 & -19.0 & 0.09 & 0.07 & 0.11 & 0.06 & 0.04 & 0.14\\
$27.6\le J\le 28.6$ & 32 & -18.5 & -19.0 & -18.0 & 0.08 & 0.04 & 0.10 & 0.04 & 0.02 & 1.00\\
\hline
$26.6\le J\le 28.6$ & 68 & -19.0 & -20.0 & -18.0 & 0.09 & 0.08 & 0.10 & 0.06 & 0.04 & 0.10\\
\hline
\end{tabular}
\\
$N_{cand}$ is the number of galaxy candidates at $z\sim 7$ used in this work,
with observed magnitudes $25.6\le J\le 26.6$,
$26.6\le J\le 27.6$, and $27.6\le J\le 28.6$ in the
six fields analysed here.
The columns $M_{UV}^{min}$ and $M_{UV}^{max}$ refer to the minimum and maximum
in absolute magnitudes of the observed sample, while $Rh_{mode}^{low}$ and
$Rh_{mode}^{up}$ indicate the 1$\sigma$ lower and upper limits for the
$Rh_{mode}$ parameters, in arcsec. The last two columns indicate the
1$\sigma$ uncertainties for the parameter $\sigma_{Rh}$, in arcsec.
\end{table*}

Overplotting simple power-law relations in Fig.\ref{sizelum}, it is
clear that there is a strong dependence of the size on 
luminosity, $Rh\propto L^{1/2}$ or $Rh\propto L^{1/3}$. We cannot
exclude however that the size distribution is a constant function
($Rh\sim 0.16$ arcsec) at $L>L^*$ and then there is a cutoff
to 0.08 arcsec (=0.4 kpc at z=7) for fainter galaxies, rather than being
represented by a smoother trend with luminosity.
The point at $M_{UV}=-18.5$ deserves also particular attention: it is the
combination of two fields in particular, the HUDF and P12HUDF, since the
five GDS galaxies at $J\sim 27.6$ are not able to provide strong constraints
to the size distribution. If we check the results of the two ultradeep fields
separately, we find two different best fit values, $Rh_{mode}=0.09\pm 0.02$
arcsec for HUDF and $Rh_{mode}=0.03\pm 0.03$ for P12HUDF.
The differences between these two samples are also evident in
Fig.\ref{ALLmag_rh}, where the typical galaxies in the HUDF have a larger
observed $Rh$ with respect to the ones in the parallel field.
The two inconsistent results can be due to the small volumes covered by
these two HST pointed observations, which can be affected by a strong field
to field variation. Larger areas covered by HST at a similar depth of the
HUDF are thus required in order to solve the inconsistency of $Rh$ estimate
at faint magnitude limits.

The faint bin of the size-luminosity relation in Fig.\ref{sizelum} is
particularly problematic also for another reason. From
Fig.\ref{rhl} it is clear that in the HUDF and P12HUDF, for galaxies
with input $Rh<0.1$, the size of galaxies as measured by SExtractor is
completely dominated by the PSF of HST and the resulting $Rh(output)$
is around 0.12 arcsec. This feature in the HUDF and P12HUDF fields is
present both for relatively bright ($J\sim 26.5$) and for faint
($J\sim 28.5$) simulated galaxies. Moreover, at $J\sim 28-28.5$, the
galaxy size is under-estimated for simulated objects with input $Rh$
greater than 0.2 arcsec. Thus, it is reasonable to think that the
present $z\sim 7$ galaxies at
$J\sim 28$ cannot give strong constraint to $Rh_{mode}$ estimation,
and the behavior of the size-luminosity relation for
$M_{UV}\ge -19$ is presently not robust.

To check the reliability of the $Rh_{mode}$ determination at the
faintest magnitudes, we have also investigated in the HUDF the power
of the maximum likelihood test on the discrimination between different
best fit solutions. The relatively large PSF of WFC3 with respect to
the expected size of galaxies at z=7 and $M_{UV}>-19$ can induce the
reader to think that it is not possible to distinguish between values
of $Rh_{mode}$ smaller that $\sim 0.1$ arcsec at $J\sim 28$. We have
verified for HUDF galaxies in the magnitude range $27.6<J<28.6$ that an
input distribution with $Rh_{mode}=0.02$ arcsec has a likelihood
parameter that is 3$\sigma$ off from the best fit ($Rh_{mode}=0.08$
arcsec), while if we choose $Rh_{mode}=0.14$ arcsec the likelihood
test can reject this solution at 2$\sigma$ level. Thus, we can
conclude that our simulations are able to distinguish between values
of $Rh_{mode}$ smaller than the actual size of the WFC3 PSF. As we
discussed above, the main uncertainty on the size determination at
$M_{UV}>-19$ is the field-to-field variance, with significantly
different values between the HUDF ($Rh_{mode}=0.09\pm 0.02$ arcsec)
and the P12HUDF field ($Rh_{mode}=0.03\pm 0.03$ arcsec).

From these results we can draw another conclusion. While we could
not exclude the presence of a non negligible population of faint and
extended galaxies, it is clear that the typical (mode) physical
dimension at $L\le L^*$ is smaller than for more luminous galaxies. A
similar result has been found on the HUDF also by
\cite{oesch09b}.

We thus have found evidence for a size-luminosity
relation (in the sense that bright galaxies can be extended while
faint galaxies are always compact/small) at high redshifts, confirming
the general trend found by star forming galaxies (LBGs, LAEs) at
lower-z (\cite{nagy,bond,law}). This is also in agreement
with the works of \cite{vanzella,pentericci10,malhotra11}: they found
that LAEs at $z\sim 3-4$ are in general smaller than the LBG sample
and fainter in the UV continuum. They also showed that galaxies with
small or negative EW in Lyman-$\alpha$ span a wide range of sizes
and UV continuum luminosities,
while large EW objects tend to be very small and faint in $M_{UV}$, in
practice line emitters tend to be small galaxies, while amongst LBGs
there are both small and large galaxies.

A comparison can be carried out at this stage between the
size/luminosity relation of LBGs at $z\sim 7$ with the one of
star-forming galaxies at lower-z. \cite{papovich05} found that the
typical size of spiral galaxies with $M_{UV}\sim -21$ at $z\sim 2-3$
is $\sim 2 kpc$, slightly larger than our determination ($\sim 0.8$
kpc) at the same luminosity. We can thus infer that the
size-luminosity relation of star-forming galaxies is evolving slowly
from $z=7$ to $z=2$, in agreement with the $(1+z)^{-1}$ evolution of
$Rh$ with redshift found by \cite{bouwens04} at lower-z and confirmed by
\cite{hathi08} at $z\sim 5-6$ and by \cite{oesch09b} at $z\sim 7$.

\begin{figure}
\centering
\includegraphics[width=8cm]{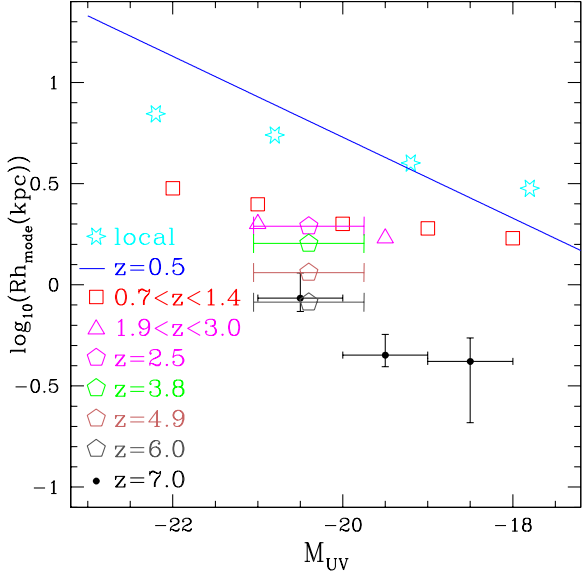}
\caption{The redshift evolution of the size-luminosity relation from
$z\sim 0$ to z=7. The cyan empty stars show the local relation
found by \cite{dejong00}, taking the absolute magnitudes in the I band
from their work and applying a constant shift $M_{UV}-M_{I}=1.0$.
The blue line is the relation by \cite{roche1996} for irregular galaxies
at $z\sim 0.5$, the empty squares and triangles represent galaxies at
$0.7<z<1.4$ and $1.9<z<3.0$ by \cite{papovich05} and the pentagons are
the typical half light radii of LBGs at z=2.5 (magenta, top), 3.8 (green),
4.9 (brown), and 6.0 (grey, bottom) by
\cite{bouwens04}. The black dots with error bars are the z=7 relation
found in this work.}
\label{evolsizelum}
\end{figure}

Local irregulars and spirals are characterized by a similar
size-luminosity relation. \cite{roche1996} derived a relation
between the half light radius and the luminosity at $z\sim 0.5$ of
$Rh\propto L^{1/2}$.
Fig.\ref{evolsizelum} summarizes the size-luminosity relations at various
redshifts described above, showing a continuous evolution in $Rh$ from z=7
(our points) to z=1-3 (the empty squares and triangles by \cite{papovich05}).
At lower redshift the situation is currently not clear, with \cite{roche1996}
finding a steep relation between the half light radius and the
luminosity of galaxies at z=0.5, $Rh\propto L^{1/2}$, while \cite{dejong00}
measured a flatter relation $Rh\propto L^{1/3}$, and slightly lower in
normalization at $M_{UV}=-21$ for the local galaxies.
Despite this discrepancy at lower-z, the normalization of the local
size-luminosity relation is higher at low-z than at $z\sim 2-7$: a
spiral/irregular galaxy with $M_{UV}\sim -21$
has a physical (proper) size of 6-10 kpc at $z<1$, compared to 2 kpc
at $z\sim 2$.
This implies that a significant evolution on the size/luminosity relation
happened in the last 8 Gyr of the lifetime of the Universe, in contrast with
a slow growth at $z\ge 2$.

\section{Discussion}

\subsection{The impact on the reionization process}

The presence of a size/luminosity relation at $z\sim 7$ has important
implications for the luminosity function of LBGs at $z\sim 7$ and the
role of stars on the reionization of the Universe. The observed lack
of extended galaxies at faint magnitude limits ($Rh\sim 0.2$ arcsec at
$J>27$) implies that the typical half light radius is less than
$\sim$0.1 arcsec ($Rh_{mode}$).

\begin{figure}
\centering
\includegraphics[width=8cm]{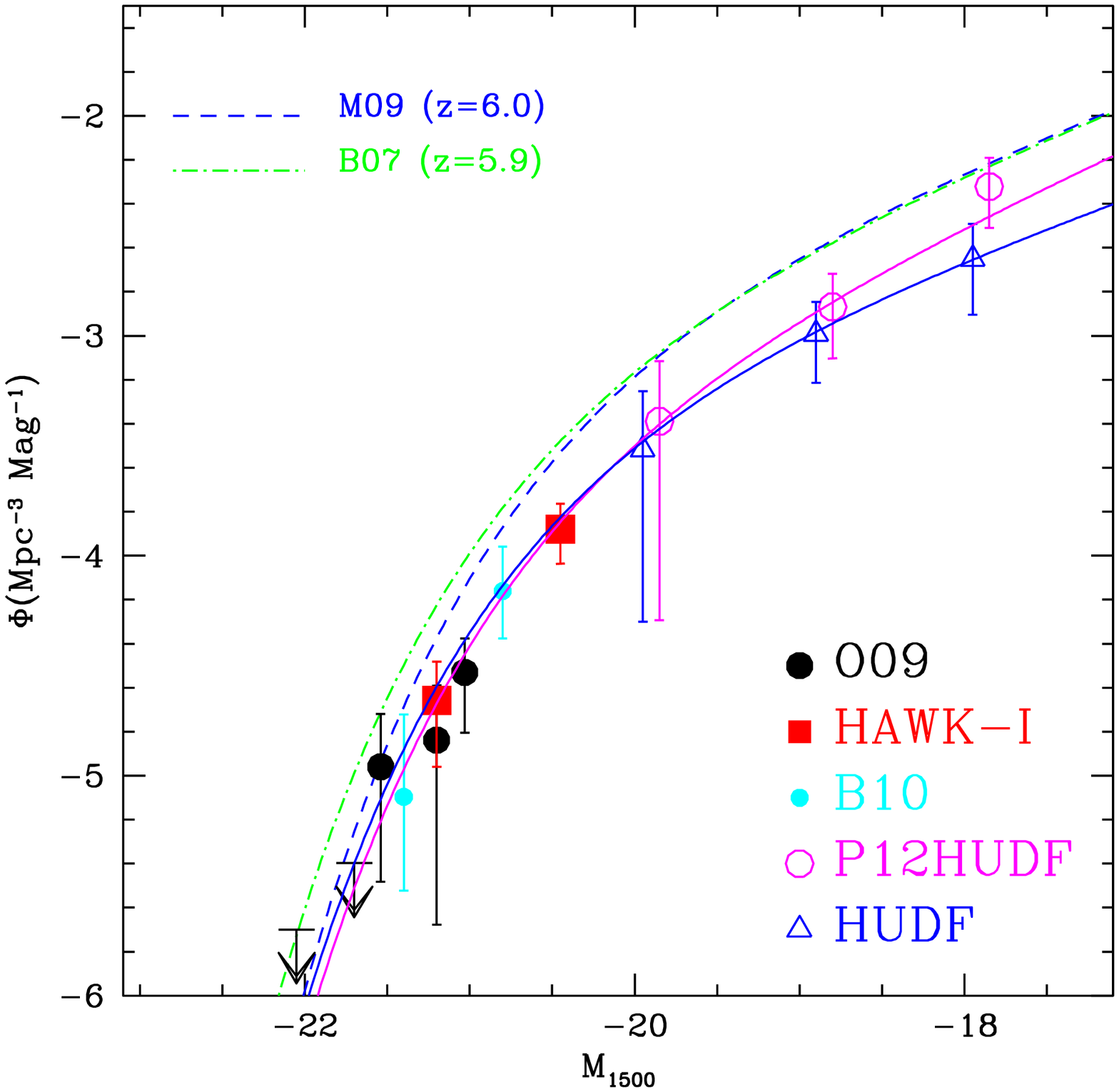}
\caption{The Luminosity Function of LBG galaxies at $z\sim 7$ computed
with the Stepwise method and assuming a log-normal size distribution with
parameters $Rh_{mode}=0.09\pm 0.01$ arcsec and $\sigma_{Rh}=0.06\pm 0.04$.
The filled symbols show the data points presented
in \cite{grazian11} (black dots and upper limits by \cite{ouchi}; red
squares by \cite{castellano10} and cyan dots by \cite{bouwens10nic}),
while the open points are the number densities for the HUDF
(triangles) and P12HUDF (circles). The cosmic variance between the two fields
is greater than the statistical uncertainties.}
\label{lfcosmvar}
\end{figure}

The present data on individual fields are not able to give stringent
constraints on the amplitude of the size distribution ($\sigma_{Rh}$),
as described above. We explore here the possibility to sum up the
likelihood regions for all the fields probing the faint side of the
magnitude distribution, namely the ERS, GDS, P12HUDF and HUDF, in the
magnitude range $26.6\le J\le 28.6$. Adding together the outputs of
these four fields, we derive a best fit of $Rh_{mode}=0.09\pm 0.01$
arcsec and a $\sigma_{Rh}=0.06\pm 0.04$. We thus explore the
dependence of the slope of the faint end of the $z\sim 7$ LBG
Luminosity Function $\alpha$ on the shape of the half-light radius
distribution. We fix the parameters $Rh_{mode}$ to 0.09 and
$\sigma_{Rh}$ to 0.06, which are the best fit values found above for
the combination of all the faint fields. We compute the number density
of $z\sim 7$ LBGs using the stepwise technique as in \cite{grazian11}
for the HUDF and P12HUDF fields separately and plot the results in
Fig.\ref{lfcosmvar}. At faint absolute magnitudes, $M_{UV}\sim -18$,
the field to field variation in the number counts is comparable with the
statistical errors (a combination of Poisson noise and uncertainties
due to the conversion of observed J band magnitudes into absolute
values taking into account the output of our simulations). In
particular, in the P12HUDF field an excess of faint $z\sim 7$ galaxies
have been detected, which results in a relatively steep LF for this
field.
We have checked that this enhancement is not due to an artificial
overcorrection due to an excess of false positive rejection rate at
faint magnitudes in the P12HUDF field. The number of simulated galaxies at
$z\sim 7$ which are not recovered by our criteria in this field is 11\% and
it is similar to that found in HUDF (12\%).

We have fitted with a Schechter function (\cite{Schechter1976}) the
individual stepwise results for HUDF and P12HUDF adding
the LF determinations at $M_{UV}\le -20.5$ discussed in \cite{grazian11}
and shown in Fig.\ref{lfcosmvar} with filled symbols.
Fixing $\Phi^*$ and $M^*$ to the values provided in \cite{grazian11},
namely 0.00074 and -20.14,
we obtained $\alpha=-1.65\pm 0.09$ and -1.83$\pm$0.18 for the HUDF and
P12HUDF, respectively.
We can derive a mean value for $\alpha$ by averaging the number
counts in the HUDF and P12HUDF fields. We thus obtained $\alpha=-1.7\pm0.1$,
where the large uncertainties reflects the strong field to field variation
affecting the present data.

If we consider $Rh_{mode}=0.09$ arcsec and $\sigma_{Rh}=0.48$, we
obtain the most extended size distribution still allowed at the 95\%
confidence level by the combined likelihood regions in the magnitude
range $26.6\le J\le 28.6$. Using this distribution into the simulations,
we derived two luminosity functions at $z\sim 7$ for the HUDF and P12HUDF
fields as described above, and fitting them with a Schechter function we
obtained $\alpha=-1.87\pm 0.12$ and $\alpha=-2.05\pm 0.23$, respectively.
Combining the two fields, a best fit of $\alpha=-1.95\pm 0.15$ has been
derived.
We thus confirm the anti-correlation between the parameter
$\sigma_{Rh}$ and the resulting steepness $\alpha$ of the LF, already
found in \cite{grazian11}. In our previous paper, we found that the
faint end of the $z\sim 7$ LBG Luminosity Function $\alpha$ is
$\sim$-1.7 (see their Fig.8), when a compact morphology ($Rh\le 0.15$
arcsec) is adopted. For more extended morphologies ($Rh\sim 0.25$
arcsec) a steeper LF was derived ($\alpha\sim-2$), confirming these
results.

A plausible value for the relevant UV emissivity of LBGs at $z=7$,
$\rho_{UV}$, can be computed by integrating the present z=7 LF
with $\alpha=-1.7$ down to $M_{1500}=-10$, assuming
that the steepness of the faint end of the LF remains constant down to
fluxes significantly fainter than that reached by our deepest
images, the HUDF and P12HUDF ones.

Using $M^*=-20.14$, $Log\Phi^*=-3.13$ for z=7 as in \cite{grazian11},
we obtain a luminosity density of $\rho_{UV}(-10)=1.1\cdot 10^{+26}$
$erg~s^{-1}~Hz^{-1}~Mpc^{-3}$ corresponding to $\alpha=-1.7$ (the 1 $\sigma$ lower
and upper bounds for this quantity are $8.1\cdot 10^{+25}$ and $1.5\cdot 10^{+26}$
$erg~s^{-1}~Hz^{-1}~Mpc^{-3}$, corresponding to $\alpha=-1.6$ and $\alpha=-1.8$,
respectively). We have no information on the number
density of $z\sim 7$ galaxies at these faint magnitudes ($M_{UV}=-10$)
with the present data, so it is useful to stress that this is a very
strong extrapolation.

Given these confidence limits on $\rho_{UV}$ and adopting the same
assumptions of \cite{grazian11}, in order
to have the Universe ionized at $z=7$ we derive a lower limit for the Lyman
Continuum escape fraction $f_{esc}\ge 0.14\times C_{HII}$ for $\alpha=-1.7$,
where $C_{HII}$ is the clumpiness of the IGM at z=7. Taking into
account the uncertainties on the faint end of the LF, this translates
into $f_{esc}\ge 0.10\times C_{HII}$ for $\alpha=-1.8$ and $f_{esc}\ge
0.19\times C_{HII}$ for $\alpha=-1.6$ (at 68\% c.l.), respectively.

In the following, we will consider the implications of a luminosity
function at z=7 with $\alpha=-1.8$, in order to relax the constraints
on the other two parameters, $f_{esc}$ and $C_{HII}$.
Assuming a maximum escape fraction of 1.0, the above limits can be
converted into constraints to the IGM at z=7, $C_{HII}\le 10.0$,
provided that the only source of UV photons are stars
and the Universe is fully ionized at z=7. Of course, smaller values
for the clumpiness $C_{HII}$ are required if the escape fraction is
less than 100\% or the LF is flatter than $\alpha=-1.8$. We must
consider $C_{HII}\ge 1$, since galaxies at $z\sim 7$ are formed in
biased density regions of the Universe and the IGM is not homogeneous
($C_{HII}=1$) at these redshifts. Limiting the clumpiness to
$C_{HII}\ge 1$, we thus have $f_{esc}\ge 0.10$ in order to have the
Universe reionized by z=7. Bolton \& Haehnelt (2007) have inferred
$C_{HII}\le 3$ at $z\sim 6$ from the Lyman-$\alpha$ forest
photoionization state: since $C_{HII}$ is expected to monotonically
decrease towards high-z in a hierarchical Universe, an escape fraction
of $f_{esc}\sim 0.3$ is enough for stellar ionizers to reach the
reionization of the IGM at z=7. A similar result has been derived by
\cite{shull2011}, based on hydrodynamical and N-body simulations, of
$C_{HII}\sim 1.5-3.0$ at $z\ge 6$. Their model has more free
parameters (the electron temperature of the IGM, the IMF of the
stellar population producing UV photons) than considered here, and
they derived that an escape fraction of 20\% is enough at z=7 to keep
the Universe ionized.

It is worth noting that the recent
estimates of $f_{esc}$ for $L\ge L^\ast$ LBGs at $0\le z\le 3$ are
$\le 10\%$ (\cite{bridge10,cowie10,siana10,vanzella10,boutsia}): thus,
assuming that the Universe is only re-ionized at $z\sim 7$ by stars in
galaxies, this implies a fast increase of the galaxy escape fraction going
to faint luminosities or to high redshifts (but see \cite{nestor} for
a larger estimate of the escape fraction of $L^\ast$ LBGs at $z\sim
3$).

We neglect in our computation the contribution of AGNs, since their LF
at z=7 is still unknown and the upper limits currently available
indicate that the AGN will add only 5-8\% to the luminosity density of
galaxies (\cite{ouchi,cowie10,haardt11}). These estimates however are derived
extrapolating the behavior of very bright QSOs selected by SDSS
($M_{UV}\sim -26$) down to very faint magnitude limits. Recent results
using the 4 Msec Chandra observation in the GOODS-South region pointed
out that the X-ray selected AGNs present a rather steep LF toward
faint magnitudes (\cite{fiore11}), they are characterized by a
significant escape fraction (half of them have $f_{esc}\sim$100\%,
see \cite{vanzella10}) and thus they could be the main
responsibles of the reionization process at $z\ge 6$.

Summarizing, we have found here that the number density of faint LBGs
at $z\sim 7$ can be fitted with a Schechter luminosity function with a
faint end slope of $\alpha=-1.7\pm 0.1$, assuming a size distribution
with $Rh_{mode}=0.09\pm 0.01$ arcsec and $\sigma_{Rh}=0.06\pm 0.04$.
The value of the parameter $\alpha$
depends critically on the size distribution of faint LBGs,
as we already found in \cite{grazian11}, and the slope of the LF at z=7 can
be steeper if the half-light radii of LBGs at this redshift
extend towards larger values.
We have detected the large field-to-field variation in
the number density of faint galaxies at $z\sim 7$, but this does not
prevent us from deriving a limit on the steepness of the z-dropout LF,
$\alpha=-1.7\pm 0.1$ at 68\% confidence level. Using this limit,
the Universe can be reionized by galaxies at large redshifts, only if
their escape fraction is larger than $\sim 30\%$.

A plausible conclusion is that the number of faint galaxies at
$z\sim 7$ in the Universe is enough to re-ionize the Universe
($\alpha\sim -1.7$) only when extrapolating the present LF down to
$M_{1500}=-10$, assuming a combination of small clumpiness for
the IGM and relatively high escape fraction of Lyman continuum
photons ($f_{esc}\ge 0.14\times C_{HII}$). A similar conclusion has
been reached also by \cite{finkelstein12}.
Since all these conditions
are extreme assumptions, we start posing some doubts that the galaxies
alone at $z\sim 7$ are able to keep the Universe ionized without the
additional contributions of faint AGNs (\cite{fiore11}) or other more
exotic explanations (\cite{dopita,conroy12}). Of course, invoking a dramatic
increase of the escape fraction with redshift $f_{esc}\propto
(1+z)^{3.4}$ as done in \cite{haardt11} could be an alternative
solution to alleviate this problem.

\subsection{Is this picture conclusive ?}

The picture of the size-luminosity relation sketched in this work
might not be conclusive, for a number of reasons: our results are
based on photometrically selected candidates and they could be
contaminated by lower-z interlopers; the bright side of the
distribution is based only on a single field (UDS) and only half of
the EGS region, and it is probably affected by the cosmic variance
effect, which should be stronger for more luminous galaxies; the
morphology adopted in this work (disk galaxies with exponential
profile) cannot be representative of the real galaxy shapes,
especially at $z\sim 7$ in the UV rest-frame, where the star-forming
galaxies are clumpy and irregular; the distribution for the axial
ratio parameter $b/a$ assumed in this work (flat from 0 to 1) may not
be representative of the real one (see \cite{ferguson04});

While we acknowledge that our analysis is not decisive, we tend to believe
that these results are robust for a number of reasons.

All the six fields used in this work
are characterized by a combination of deep multi-wavelength photometry,
which ensures a clean sample of candidate galaxies at $z\sim 7$.
The spectroscopic confirmations of the bright candidates in the GOODS
fields, and in other ground based surveys, is currently ongoing
(\cite{fontana10,stark,vanzella11,ono}) and the fraction of lower-z
interlopers seems to be less than 20\% at $z\sim 6$ (\cite{pentericci11}).
Thus we are confident that the sample adopted for studying the size-luminosity
relation in this paper is not heavily contaminated by lower-z galaxies.

At the moment, the bright side of the $z\sim 7$ galaxy population is
sampled mainly by the UDS field, so the robustness of the
size-luminosity relation is based only on this region. Since the
cosmic variance effect should be stronger for more luminous galaxies,
one can suspect that it is only by chance that bright galaxies in the
UDS are also the largest, when compared to the ones in deeper fields
(i.e. HUDF). However, at $J\le 25.8$ we have 7 galaxies with $Rh\ge
0.2 arcsec$ out of a total of 14 objects (50\%) in the combined UDS
and EGS fields, which is hard to be interpreted only as a simple
cosmic variance effect.
In the near future the
CANDELS-Wide survey will complete the EGS and cover the COSMOS field
at a similar depth and areal coverage of the UDS, thus enhancing by a
factor of two the number of galaxies in the bright side of the
size-luminosity relation and reducing the uncertainties on
$Rh_{mode}$.

The size-luminosity relation measured in this work could also be an
artifact due to the frequent merging of galaxies expected at very
high-z. Instead of measuring a physical size, the large $Rh$ found
for relatively bright objects could be an estimate for the separation
of their clumps during the merging phase.
Detailed kinematic
analysis with IFU spectroscopy at Extremely Large Telescopes (ELTs) of
30-40 meters of diameter, as done currently with 8m class telescopes
on $z\sim 2$ galaxies (\cite{forster,law}), would distinguish these
two plausible hypotheses.

A more reliable distribution for the axial ratio parameter $b/a$,
following \cite{ferguson04} can be adopted in our simulations, but we
must stress that, in their work, this functional form is measured on a
sample of
photometrically selected candidate galaxies at $z\sim 4$, and it is
not clear whether this should be applied also to our $z\sim 7$
sample. A more detailed approach, like that adopted in \cite{law}
work, is more appropriate but very complex and goes beyond the aims of
this paper.


\section{Conclusions}

Galaxy sizes (half light radii) have been measured for a sample of
153 galaxy candidates at $z\sim 7$ from the CANDELS HST Multi-Cycle Treasury
Program (\cite{grogin11,koekemoer11}) and HUDF09 project (\cite{bouwens10c}).
In particular, we have used the deep HST
imaging database in BVIZYJH bands for the ERS, GDS, P12HUDF and HUDF fields
together with the wide area observations in the UDS and EGS fields in the
VIJH bands.
We select the galaxy candidates at z=7 through the classical z-dropout
technique, which has been verified by deep VLT spectroscopy
(\cite{pentericci11}).
For the UDS and EGS we use the HST $I_{814}$ band as dropout to select high-z
galaxy candidates.

Despite the difficulties of measuring galaxy morphology at $z\sim 7$,
thanks to detailed and extensive simulations, we
successfully detect a clear size-luminosity relation for LBGs at high-z.
In particular, we found evidences that:
\begin{itemize}
\item
{\bf bright galaxies can be large.} At magnitude brighter than
$J=26.6$ (corresponding $\sim L^\ast$ at z=7) galaxies have been observed
at larger dimensions ($Rh\sim 0.4 arcsec$ or equivalently 2.3 kpc proper)
than at faint magnitudes. Again, Fig.\ref{ALLmag_rh} shows the extended
tail in $Rh$ which is present only for bright galaxies: despite all
the deeper fields are sensitive to such extended galaxies, none of
them have been found at $J\ge 26.6$.
\item
{\bf faint galaxies are small.} At $J\ge 26.6$ the observed sizes of
z=7 galaxies are smaller than 0.2 arcsec (corresponding to 1.15 kpc
proper). This is evident looking both at Fig.\ref{ALLmag_rh} and as a 
result of the detailed simulations summarized by Table \ref{tab:resu}.
\item
{\bf a size-luminosity relation is already in place at z=7.}
The observed dependency of LBG sizes from galaxy luminosity at z=7 has been
shown in Fig.\ref{sizelum}. We have found that a relation exist between these
two observables, $Rh\propto L^\gamma$, with $\gamma\sim 1/3-1/2$.
\end{itemize}

These results have deep implications for our understanding of the
reionization of the Universe. The derived size-luminosity relation at
$z\sim 7$ and the fact that faint LBGs have typical half light radii
of $\sim$0.1 arcsec seems to indicate that the slope of the z=7 LF is
not extremely steep, due to the correlation between size and $\alpha$
found in \cite{grazian11} and recovered here. We detect also a strong
field to field variation in the faint regime, with $\alpha\sim -1.6$
in the HUDF and $\alpha\sim -1.8$ in the P12HUDF field.
Using an average value for the two samples, we derived
$\alpha=-1.7\pm 0.1$.
The relevant UV emissivity of LBGs
at $z=7$, $\rho_{UV}$, has been computed by integrating the best fit
LF down to $M_{1500}=-10$, and resulted in $\rho_{UV}=1.1\cdot 10^{+26}$
$erg~s^{-1}~Hz^{-1}~Mpc^{-3}$. This amount of radiation is not able to keep the
Universe re-ionized if the IGM is clumpy ($C_{HII}\ge 3$) and if the
Lyman continuum escape fraction of high-z LBGs is relatively low
($f_{esc}\le 0.3$). The only configuration that allows a non-neutral
Universe due to stellar ionizers in galaxies is if the LBG LF is
steeper than $-1.7$, combined with a small
clumpiness for the IGM and a high escape fraction
of Lyman continuum photons. These are not implausible
conditions (see \cite{Bolton2007} or \cite{haardt11}), but of course they are
extreme assumptions, that could be overcome by simpler explanations,
like an additional contribution of faint AGNs (see \cite{fiore11}),
or more exotic explanations (see \cite{dopita}).

Our results on the size distribution and size-luminosity relation of
$z=7$ LBGs have been investigated through detailed and realistic
simulations and thus they are robust conclusions. However, the dataset
used, especially at the bright side (UDS), is not free from biases due
to cosmic variance effects, and we cannot exclude that subtle effects
can modify our results. Investigating wide fields searching for
$-21\le M_{UV}\le -20$ galaxies will
be very useful to reinforce our statements. The CANDELS-Wide survey,
by observing the COSMOS (\cite{cosmos}) and the whole EGS (\cite{egs}) fields,
for a total of other 300 sq. arcmin down to J=26.7, will provide $\sim
70$ additional bright candidates at $z=7$, and would be able to beat
down the cosmic variance affecting the bright side of $z=7$
distribution. In addition, WFC3 imaging of $Y\sim 25$ z-dropout
galaxies found with large area ground based imaging
(\cite{ouchi,castellano09,castellano10,bowler12})
will provide useful information on
the bright side of the size-luminosity relation not yet covered by the
present observations.

Moreover, the CANDELS Deep survey plus the HUDF ultradeep
fields (both those already observed and the
HST program recently approved for Cycle 19) will extend these results and
confirm in a more accurate statistical evidence the trend of the
size-luminosity relation at $z\sim 7$. In
particular, the CANDELS-Deep region on the two GOODS fields, covering
150 sq. arcmin down to $J\sim 28$, will open a very interesting window
on the exact shape of the size distribution down to very large half
light radii. In addition, the combination of depth and area guaranteed by
the CANDELS-Deep survey will decrease the uncertainties on the faint side
determination of the $z\sim 7$ LF due to cosmic variance.

Clearly, the size-luminosity relation found here is the simplest
correlation between physical properties of galaxy at $z\sim 7$. With a
full spectroscopic sample in hand, coupled with the deep
multi-wavelength dataset available for the CANDELS survey, it will be
possible to explore the dependencies of the half light radius with the
galaxy stellar mass, SFR, dust extinction or the EW of Ly-$\alpha$, as
already done for star-forming galaxies at smaller redshifts.

\begin{acknowledgements}
We thank the referee for her/his useful comments and suggestions that
helped us to improve the paper.
We acknowledge financial contribution from the agreement ASI-INAF I/009/10/0.
EV acknowledges financial contribution from PRIN
MIUR 2009 ``Tracing the growth of structures in the Universe:
from the high-redshift cosmic web to galaxy clusters''.
This work is based on observations taken by the CANDELS Multi-Cycle
Treasury Program with the NASA/ESA HST, which is operated by the
Association of Universities for Research in Astronomy, Inc., under
NASA contract NAS5-26555.
Observations were also carried out using the Very Large Telescope at
the ESO Paranal Observatory under Programme IDs LP181.A-0717,
LP168.A-0485, ID 170.A-0788, ID 181.A-0485, ID 283.A-5052 and the ESO Science
Archive under Programme IDs 67.A-0249, 71.A-0584, 73.A-0564, 68.A-0563,
69.A-0539, 70.A-0048, 64.O-0643, 66.A-0572, 68.A-0544, 164.O-0561,
163.N-0210, and 60.A-9120.
\end{acknowledgements}

%

\Online

\begin{appendix} 

\section{$z\sim 7$ galaxy candidates in the GDS, P12HUDF, EGS and UDS fields}

The $z\sim 7$ candidate galaxies in the ERS and HUDF fields have been already
presented in \cite{grazian11}. Here we provide the lists of the z-band
drop-out candidates for the GDS (Table \ref{tab:gds}), P12HUDF
(Table \ref{tab:p12hudf}), and I814-band dropout from the UDS
(Table \ref{tab:uds}) and EGS (Table \ref{tab:egs}) fields.
In Table \ref{tab:uds} (\ref{tab:egs}) we mark the 27 (13) candidates
selected by
\cite{mclure12} with a robust photometric redshift consistent with $z\ge 7$
in the UDS and EGS fields, respectively.

\begin{table}
\caption{$z\sim 7$ galaxy candidates in the GDS field}
\label{tab:gds}
\centering
\begin{tabular}{l c c c c}
\hline\hline
ID & RAD & DEC & J & Rh(arcsec) \\
\hline
565   & 53.074491 & -27.877560 & 26.77 & 0.167 \\
959   & 53.079016 & -27.873006 & 27.07 & 0.187 \\
1240  & 53.072728 & -27.869255 & 26.48 & 0.258 \\
1710  & 53.053794 & -27.864331 & 27.33 & 0.175 \\
1903  & 53.101737 & -27.862297 & 26.50 & 0.116 \\
2197  & 53.086466 & -27.859149 & 26.99 & 0.140 \\
2610  & 53.123502 & -27.855121 & 27.67 & 0.199 \\
2613  & 53.112600 & -27.855104 & 27.15 & 0.264 \\
2721  & 53.091561 & -27.853837 & 26.66 & 0.204 \\
3331  & 53.106017 & -27.848158 & 26.54 & 0.161 \\
6444  & 53.101396 & -27.820836 & 27.64 & 0.131 \\
6483  & 53.099068 & -27.820457 & 26.17 & 0.321 \\
6619  & 53.036200 & -27.819195 & 27.64 & 0.142 \\
7015  & 53.064118 & -27.815569 & 27.33 & 0.160 \\
7817  & 53.058939 & -27.808653 & 27.60 & 0.189 \\
8899  & 53.066929 & -27.799365 & 26.95 & 0.215 \\
9705  & 53.052408 & -27.791223 & 25.73 & 0.149 \\
10362 & 53.096474 & -27.787018 & 26.99 & 0.134 \\
10453 & 53.083067 & -27.786274 & 27.80 & 0.134 \\
11438 & 53.033874 & -27.778009 & 26.32 & 0.188 \\
13233 & 53.083367 & -27.764388 & 27.28 & 0.119 \\
\hline
\end{tabular}
\end{table}

\begin{table}
\caption{$z\sim 7$ galaxy candidates in the P12HUDF field}
\label{tab:p12hudf}
\centering
\begin{tabular}{l c c c c}
\hline\hline
ID & RAD & DEC & J & Rh(arcsec) \\
\hline
173  & 53.263422 & -27.703536 & 28.50 & 0.189 \\
330  & 53.264582 & -27.700250 & 27.26 & 0.352 \\
337  & 53.258939 & -27.700140 & 28.40 & 0.140 \\
355  & 53.255364 & -27.699911 & 28.12 & 0.235 \\
686  & 53.266042 & -27.693785 & 28.60 & 0.197 \\
801  & 53.260144 & -27.692035 & 27.51 & 0.141 \\
1059 & 53.248301 & -27.689142 & 27.19 & 0.187 \\
1212 & 53.283644 & -27.687548 & 27.72 & 0.171 \\
1213 & 53.249308 & -27.687531 & 28.29 & 0.115 \\
1271 & 53.265814 & -27.686757 & 28.79 & 0.123 \\
1278 & 53.284528 & -27.686671 & 27.93 & 0.106 \\
1364 & 53.236245 & -27.685609 & 27.01 & 0.176 \\
1403 & 53.232299 & -27.685280 & 27.96 & 0.104 \\
1404 & 53.240862 & -27.685269 & 28.43 & 0.149 \\
1545 & 53.234488 & -27.683431 & 28.41 & 0.134 \\
1634 & 53.253769 & -27.682253 & 28.24 & 0.193 \\
1744 & 53.245689 & -27.680663 & 27.41 & 0.190 \\
1750 & 53.245773 & -27.680770 & 28.68 & 0.137 \\
1988 & 53.254990 & -27.677560 & 28.78 & 0.122 \\
2005 & 53.248703 & -27.676469 & 25.94 & 0.162 \\
2070 & 53.238318 & -27.676401 & 28.10 & 0.115 \\
2226 & 53.243749 & -27.673300 & 27.85 & 0.122 \\
2261 & 53.244625 & -27.672749 & 28.65 & 0.129 \\
\hline
\end{tabular}
\end{table}

\begin{table}
\caption{$z\sim 7$ galaxy candidates in the UDS field}
\label{tab:uds}
\centering
\begin{tabular}{l c c c c c}
\hline\hline
ID & RAD & DEC & J & Rh(arcsec) & zphot \\
\hline
2047  & 34.280510 & -5.2683411	& 26.59	& 0.151 & yes \\
2635  & 34.488762 & -5.2656941	& 25.72	& 0.184 & no  \\
2638  & 34.488934 & -5.2657399	& 26.64	& 0.140 & no  \\
2782  & 34.424973 & -5.2652078	& 26.54	& 0.156 & yes \\
3147  & 34.305576 & -5.2637830	& 26.43	& 0.176 & no  \\
4101  & 34.389111 & -5.2595940	& 26.49	& 0.174 & no  \\
4154  & 34.422218 & -5.2592878	& 25.92	& 0.182 & yes \\
6374  & 34.475731 & -5.2484989	& 25.62	& 0.159 & no  \\
6409  & 34.280376 & -5.2483692	& 25.70	& 0.326 & yes \\
6443  & 34.482029 & -5.2481699	& 25.66	& 0.148 & yes \\
9386  & 34.321495 & -5.2355309	& 26.23	& 0.214 & yes \\
10527 & 34.415421 & -5.2307229	& 26.02	& 0.179 & yes \\
13324 & 34.367481 & -5.2190270	& 26.18	& 0.239 & no  \\
13796 & 34.224491 & -5.2169590	& 26.04	& 0.244 & no  \\
14435 & 34.323608 & -5.2141371	& 26.57	& 0.172 & no  \\
15399 & 34.233883 & -5.2100158	& 25.43	& 0.241 & yes \\
15482 & 34.253719 & -5.2095661	& 26.00	& 0.160 & yes \\
16119 & 34.253719 & -5.2068028	& 26.30	& 0.189 & yes \\
16669 & 34.279049 & -5.2043710	& 26.48	& 0.148 & yes \\
16910 & 34.226192 & -5.2033339	& 26.50	& 0.121 & yes \\
16974 & 34.313725 & -5.2030821	& 26.02	& 0.247 & yes \\
17160 & 34.313225 & -5.2023530	& 26.37	& 0.180 & no  \\
18356 & 34.325001 & -5.1977878	& 26.59	& 0.160 & no  \\
18777 & 34.283592 & -5.1959820	& 26.34	& 0.186 & yes \\
18797 & 34.278481 & -5.1958232	& 26.16	& 0.145 & no  \\
18932 & 34.482838 & -5.1953049	& 26.17	& 0.158 & yes \\
19136 & 34.467083 & -5.1944618	& 26.25	& 0.193 & yes \\
19390 & 34.434162 & -5.1934099	& 26.10	& 0.170 & no  \\
19476 & 34.354675 & -5.1930408	& 26.03	& 0.179 & yes \\
19818 & 34.376148 & -5.1916552	& 26.55	& 0.156 & yes \\
21850 & 34.408279 & -5.1825972	& 25.53	& 0.410 & no  \\
23427 & 34.298386 & -5.1760311	& 25.83	& 0.234 & yes \\
24592 & 34.359615 & -5.1711669	& 25.58	& 0.115 & yes \\
25519 & 34.308323 & -5.1673961	& 26.65	& 0.147 & yes \\
26120 & 34.443848 & -5.1646738	& 26.21	& 0.196 & yes \\
26125 & 34.337284 & -5.1646609	& 26.35	& 0.151 & yes \\
27601 & 34.447922 & -5.1583772	& 25.86	& 0.224 & no  \\
28737 & 34.229103 & -5.1533098	& 25.97	& 0.210 & yes \\
29249 & 34.226135 & -5.1510921	& 25.98	& 0.143 & no  \\
29272 & 34.493279 & -5.1510839	& 26.12	& 0.160 & no  \\
30492 & 34.490051 & -5.1458049	& 25.79	& 0.188 & yes \\
32650 & 34.479736 & -5.1365409	& 26.61	& 0.148 & no  \\
33062 & 34.413895 & -5.1347852	& 25.86	& 0.179 & yes \\
33622 & 34.269180 & -5.1324148	& 26.28	& 0.188 & no  \\
33684 & 34.446419 & -5.1321688	& 26.70	& 0.122 & no  \\
34364 & 34.313526 & -5.1293492	& 25.68	& 0.196 & yes \\
\hline
\end{tabular}
We mark with the label $zphot=yes$ the 27 candidates
selected by
\cite{mclure12} with a robust photometric redshift consistent with $z\ge 7$.
\end{table}

\begin{table}
\caption{$z\sim 7$ galaxy candidates in the EGS field}
\label{tab:egs}
\centering
\begin{tabular}{l c c c c c}
\hline\hline
ID & RAD & DEC & J & Rh(arcsec) & zphot \\
\hline
1976  & 214.854416 & 52.759621 & 25.52 & 0.210 & no \\
4022  & 214.732269 & 52.685204 & 25.81 & 0.199 & yes \\
5970  & 214.779892 & 52.732086 & 26.26 & 0.152 & no \\
6168  & 215.091064 & 52.954372 & 26.62 & 0.149 & yes \\
6845  & 215.066467 & 52.941925 & 25.64 & 0.303 & yes \\
7908  & 215.188446 & 53.033737 & 26.25 & 0.205 & no \\
8053  & 215.145386 & 53.004257 & 25.32 & 0.196 & no \\
8371  & 214.710114 & 52.696918 & 26.62 & 0.160 & yes \\
10433 & 215.081482 & 52.972218 & 25.43 & 0.203 & yes \\
10671 & 215.060745 & 52.958767 & 26.56 & 0.162 & yes \\
10865 & 214.705551 & 52.707199 & 26.26 & 0.199 & yes \\
13537 & 214.677643 & 52.703339 & 26.22 & 0.228 & no \\
14449 & 214.946732 & 52.900570 & 26.64 & 0.159 & no \\
14516 & 214.945587 & 52.900261 & 26.13 & 0.146 & yes \\
15250 & 214.998825 & 52.942127 & 26.11 & 0.248 & yes \\
15423 & 215.130081 & 53.035568 & 26.57 & 0.141 & yes \\
15937 & 215.095123 & 53.014256 & 25.46 & 0.218 & yes \\
19048 & 214.670349 & 52.731167 & 26.40 & 0.149 & yes \\
19761 & 215.077621 & 53.026142 & 26.54 & 0.145 & no \\
20105 & 214.628082 & 52.709671 & 26.03 & 0.225 & no \\
21147 & 214.863052 & 52.889496 & 26.19 & 0.182 & yes \\
\hline
\end{tabular}
We mark with the label $zphot=yes$ the 13 candidates
selected by
\cite{mclure12} with a robust photometric redshift consistent with $z\ge 7$.
\end{table}

\newpage

\section{Output of completeness simulations in the EGS, ERS, GDS, P12HUDF,
and HUDF fields}

\begin{figure}
\includegraphics[width=8cm]{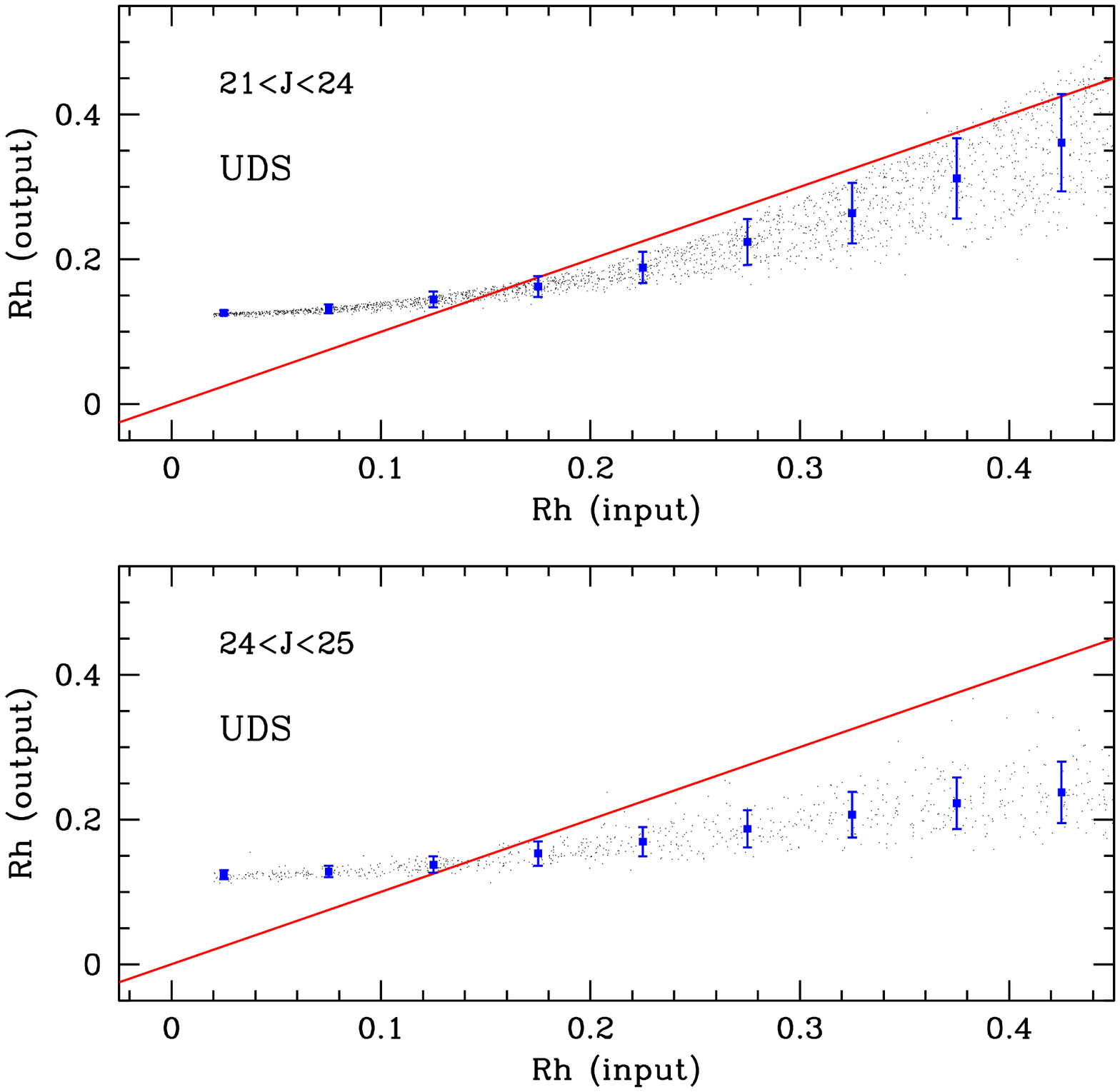}
\caption{The results of our simulations for $z\sim 7$ galaxies with Sersic
profiles of index $n=4$ in the UDS field.
The plot shows the comparison between the half--light radius (in arcsec) as
measured by SExtractor as a function of the input one. The upper and
lower panel refer to different total magnitudes, as reported in the legend.
The red line shows the
identity relation. The blue points and errorbar shows the average
value and the relevant r.m.s. of the output half--light radius.
At small sizes the output half light radius is typically larger than the
input one due to the convolution with the instrumental PSF carried out
during the simulations.}
\label{rhl_udsell}
\end{figure}

\begin{figure}
\includegraphics[width=8cm]{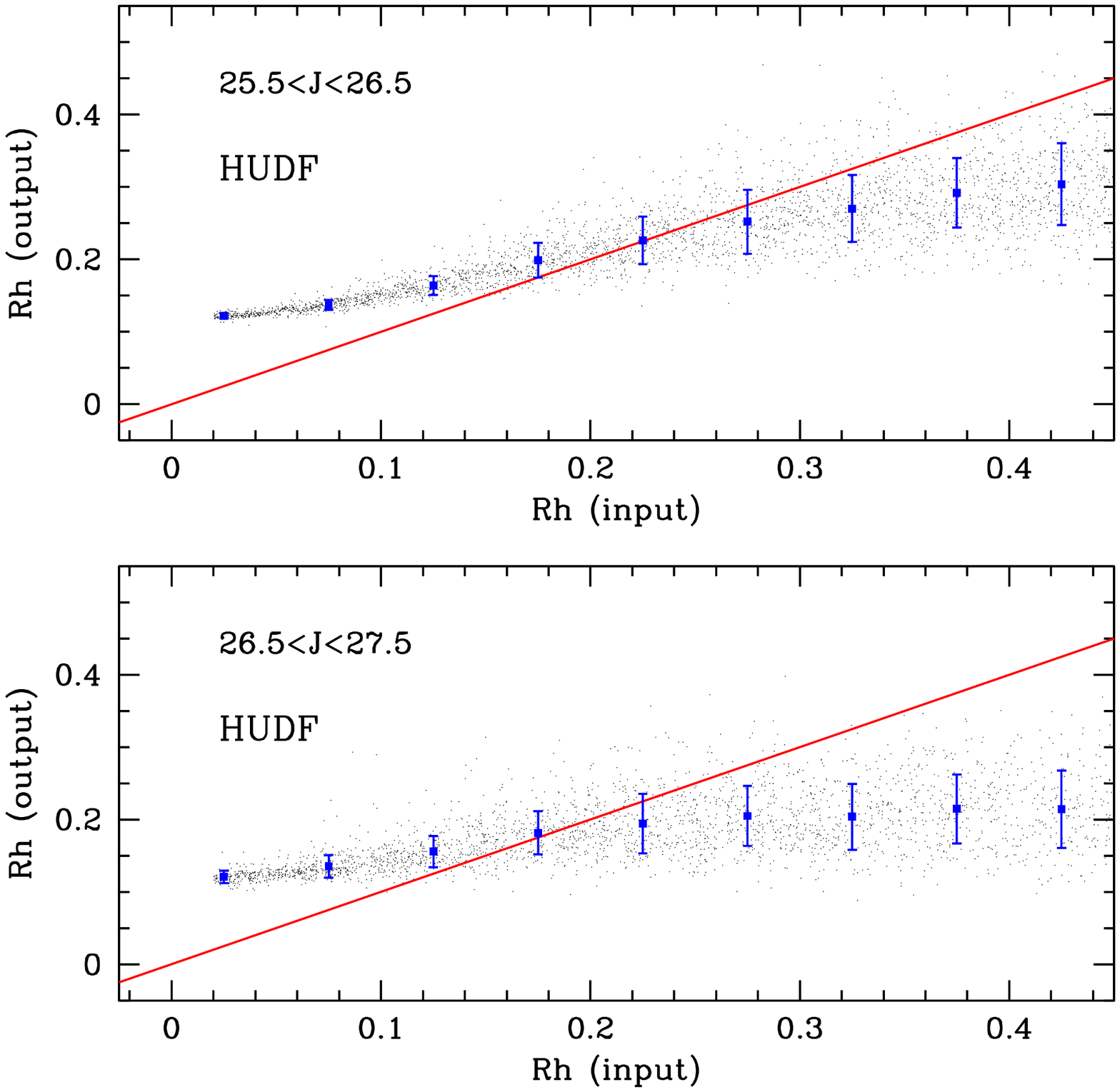}
\caption{The results of our simulations for $z\sim 7$ galaxies with Sersic
profiles of index $n=4$ in the HUDF field.
The plot shows the comparison between the half--light radius (in arcsec) as
measured by SExtractor as a function of the input one. The upper and
lower panel refer to different total magnitudes, as reported in the legend.
The red line shows the
identity relation. The blue points and errorbar shows the average
value and the relevant r.m.s. of the output half--light radius.
At small sizes the output half light radius is typically larger than the
input one due to the convolution with the instrumental PSF carried out
during the simulations.}
\label{rhl_hudfell}
\end{figure}

Fig.\ref{rhl_udsell} shows the results of the simulations carried out
in the UDS field. A Sersic profile of index $n=4$ has been adopted as
input morphology for our simulations. This plot shows the comparison
between the half--light radius (in arcsec) as measured by SExtractor
as a function of the input one, for the magnitude range $21<J<24$
(top) and $24<J<25$ (bottom). Fig.\ref{rhl_hudfell} provides the same
information for simulated galaxies in the HUDF field, in the range
$25.5<J<26.5$ (top) and $26.5<J<27.5$ (bottom). The measured size of
the objects cannot be smaller than the instrumental PSF (which is
about 0.18'' of FWHM in the J band, corresponding to an half light
radius of 0.11''). Because of the convolution with the PSF, all
objects intrinsically smaller than $\simeq 0.2$'' are biased high by
the SExtractor estimate.

\newpage

\begin{figure*}
\centering
\includegraphics[width=8cm]{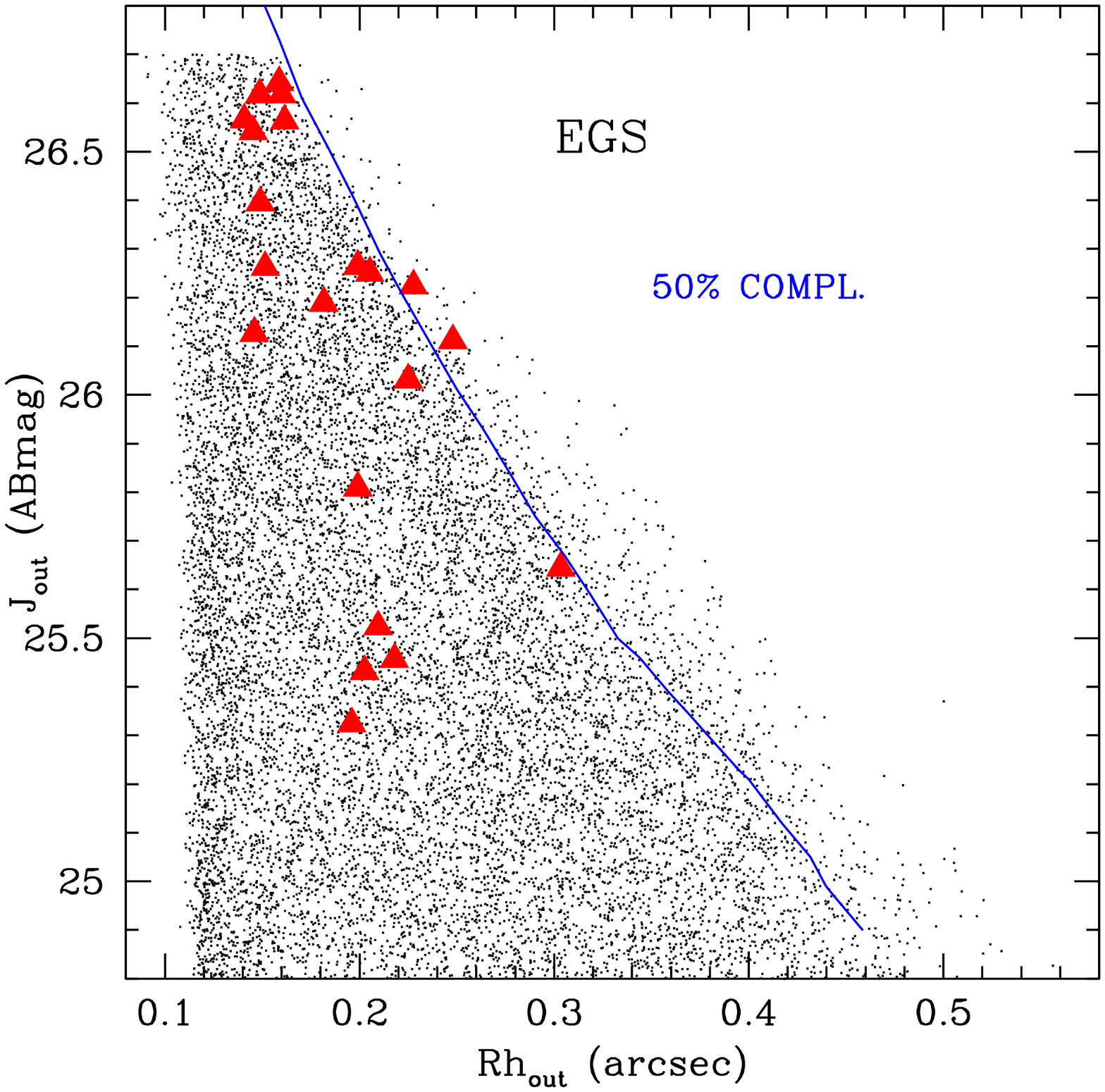}
\includegraphics[width=8cm]{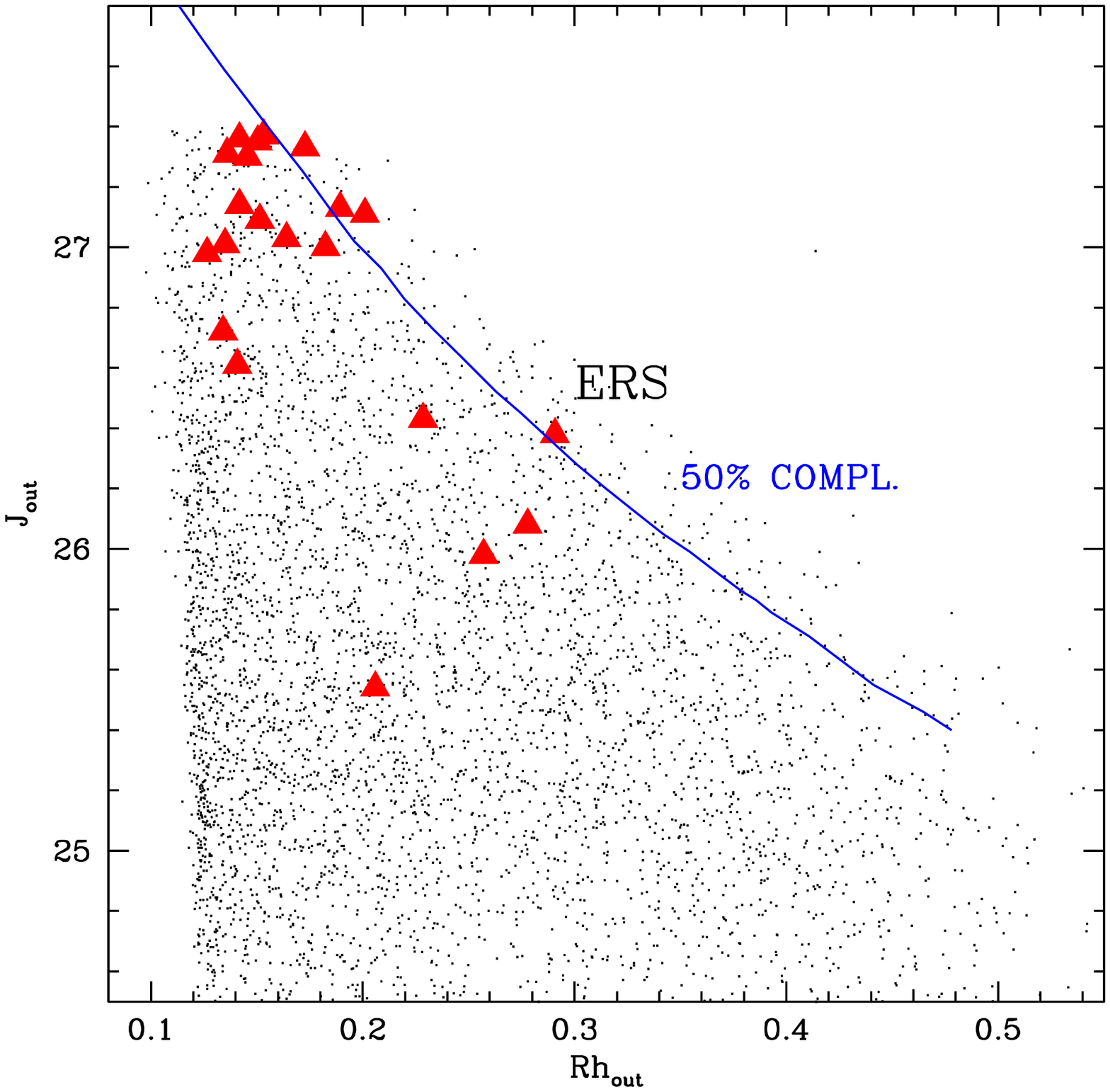}
\includegraphics[width=8cm]{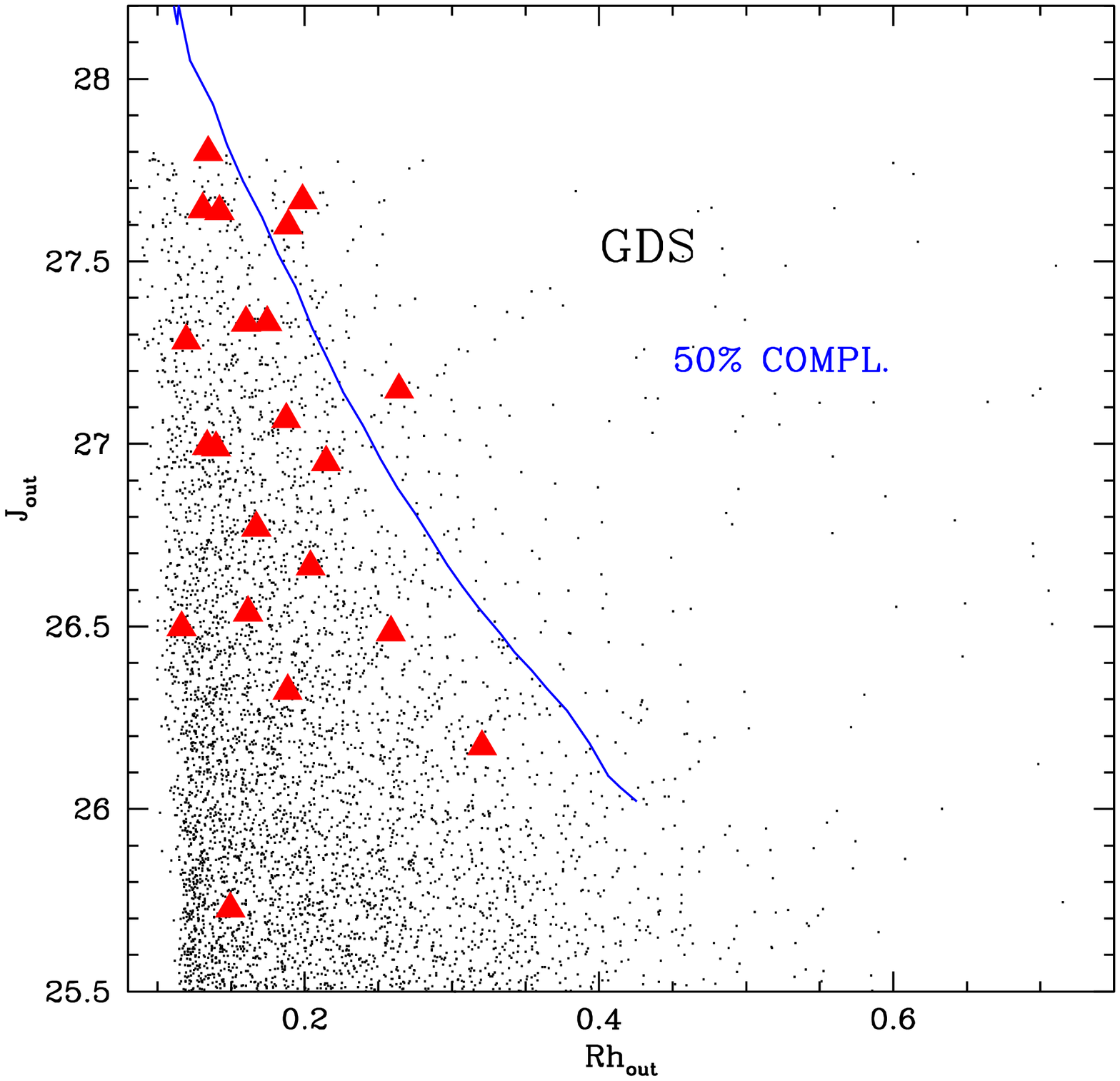}
\includegraphics[width=8cm]{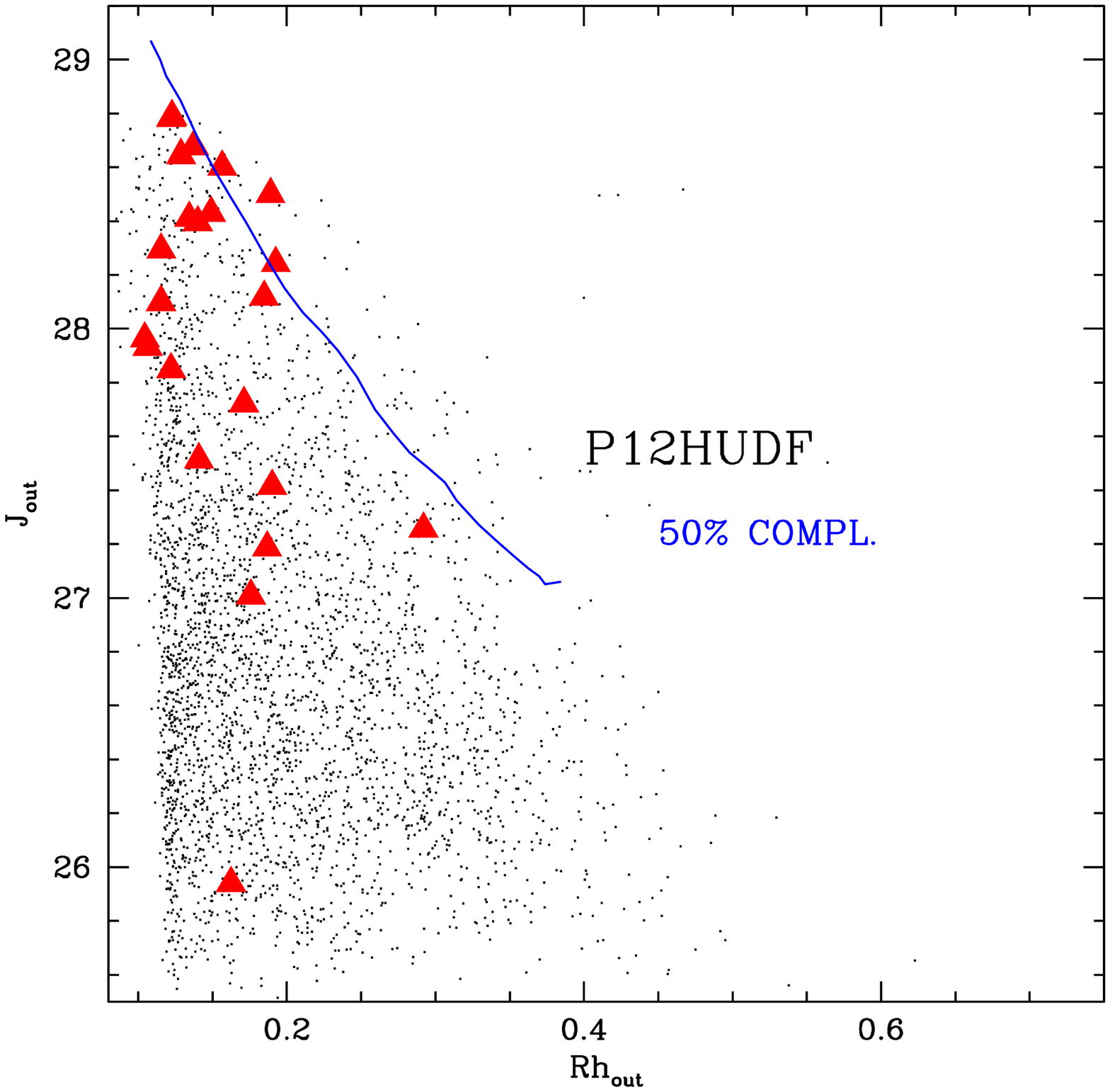}
\includegraphics[width=8cm]{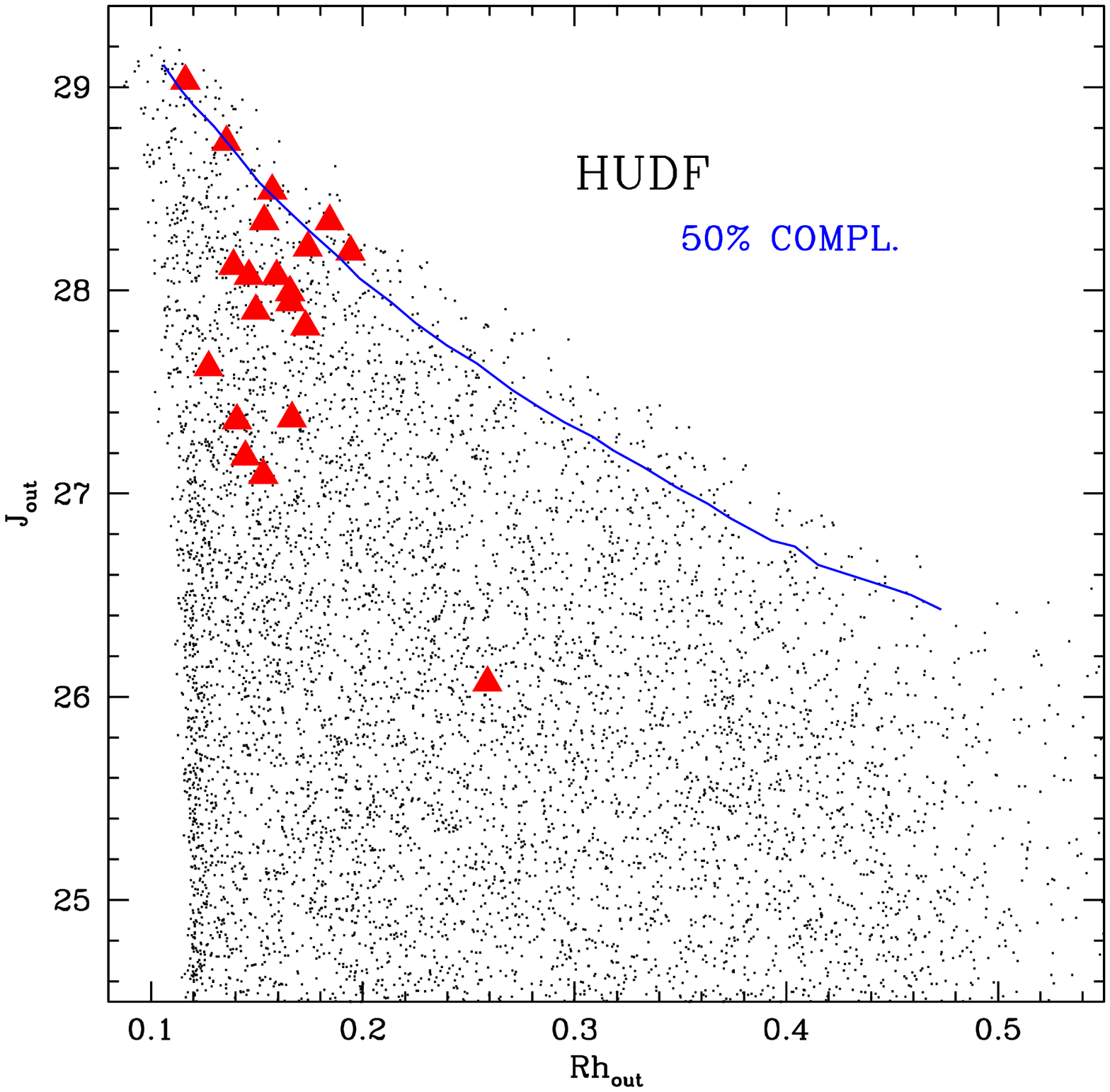}
\caption{The observed J magnitude vs size of simulated (small dots) and
observed galaxies (red triangles) at $z\sim 7$ for the EGS, ERS, GDS, P12HUDF
and HUDF fields.}
\label{FAINTmag_rh}
\end{figure*}

Fig.\ref{FAINTmag_rh} shows the observed J-band magnitudes versus the
measured sizes for simulated (small dots) and
observed galaxies (red triangles) at $z\sim 7$ for the EGS, ERS, GDS, P12HUDF
and HUDF fields. The solid blue line shows the 50\% completeness level
for an input simulated profile of disk galaxy.
Detailed explanations on the procedure
adopted in these simulations can be found in section 4.1.

\section{Best Fit}

The confidence levels for the maximum likelihood analysis of the
size distribution on the combined ERS and GDS fields (intermediate sample),
and the combined P12HUDF and HUDF fields (faint sample).
The best fit is indicated by the magenta point
while the green, blue, and red regions define the uncertainties at 68\%, 95\%,
and 99.7\% (1,2, and 3 sigma) confidence level, respectively.

\begin{figure*}
\centering
\includegraphics[width=8cm]{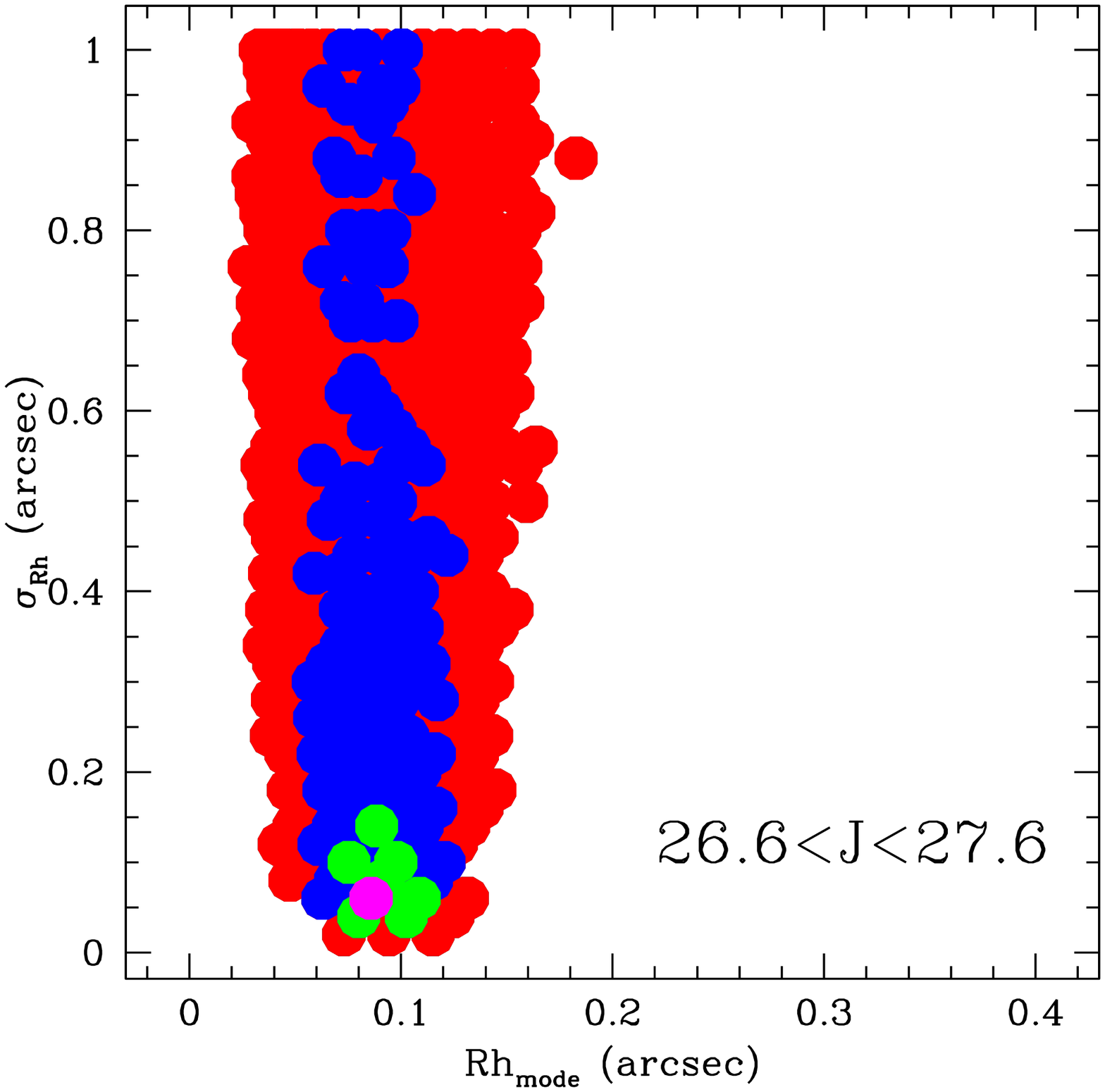}
\includegraphics[width=8cm]{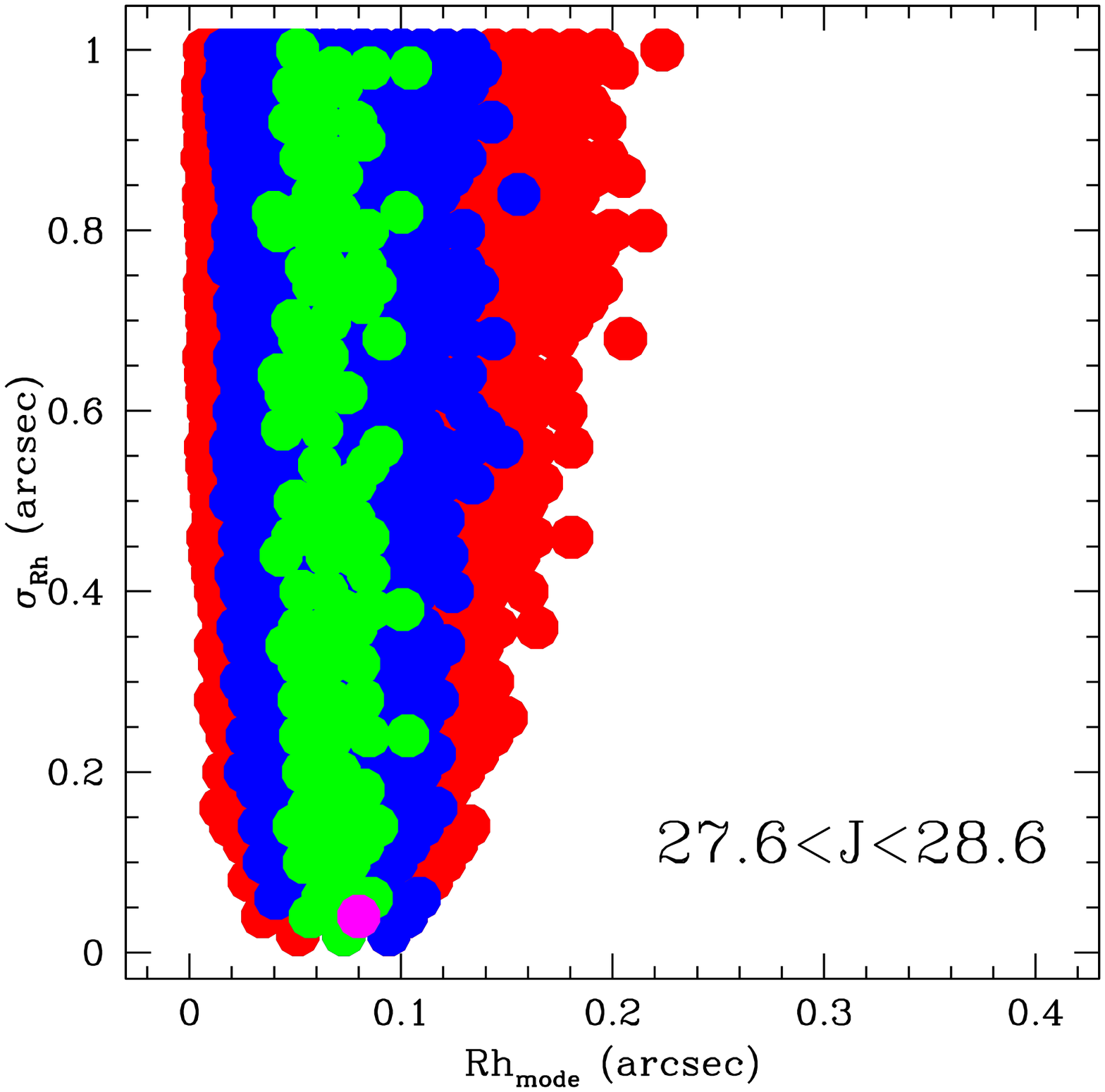}
\caption{The composite likelihood distribution for the intermediate
($26.6\le J\le 27.6$, left) and faint ($27.6\le J\le 28.6$, right)
magnitude bins. The best fit is indicated by the magenta point
while the green, blue, and red regions define the uncertainties at 68\%, 95\%,
and 99.7\% (1,2, and 3 sigma) confidence level, respectively.}
\label{regionallfaint}
\end{figure*}

\end{appendix}

\end{document}